\documentclass[12pt,letterpaper]{article}                           

\usepackage{a4}          

\usepackage{amsmath}
\usepackage{amssymb}

\usepackage{ifpdf}
\ifpdf
  \usepackage[pdftex, 
    pdftitle={Global F-theory GUTs}, 
    pdfauthor={Thomas W. Grimm, Sven Krause, and Timo Weigand}, 
    pdfsubject={String theory model building},
    bookmarksopen, bookmarksnumbered, bookmarksopenlevel=2]{hyperref}
\fi

\def\hybrid{\topmargin -20pt    \oddsidemargin 0pt
        \headheight 0pt \headsep 0pt
        \textwidth 6.25in       
        \textheight 9 in       
        \marginparwidth .875in
        \parskip 5pt plus 1pt 
          \jot = 1.5ex
   }
\hybrid
\numberwithin{equation}{section}
\numberwithin{table}{section}

\newcommand{\beq}{\begin{equation}}  \newcommand{\eeq}{\end{equation}}
\newcommand{\bal}{\begin{aligned}}   \newcommand{\eal}{\end{aligned}}
\newcommand{\bea}{\begin{eqnarray}}  \newcommand{\eea}{\end{eqnarray}}

\def\ov{\overline}

\newcommand{\ccP}{\mathcal{P}}

\newcommand{\bbC}{\mathbb{C}}
\newcommand{\bbP}{\mathbb{P}}

\newcommand{\surjto}{\to\!\!\!\!\!\to}

\newcommand{\tablelabel}[1]{\label{table:#1}}
\newcommand{\tableref}[1]{Table~\ref{table:#1}}
\newcommand{\Tableref}[1]{Table~\ref{table:#1}}

\newcommand{\seclabel}[1]{\label{sec:#1}}

\newcommand{\cL}{\mathcal{L}}

\newcommand{\cN}{\mathcal{N}}

\begin{document}

\baselineskip=14pt
\parskip 5pt plus 1pt

\vspace*{-1.5cm}
\begin{flushright}    
  {\small
  HD-THEP-09-30\\
  }
\end{flushright}

\vspace{2cm}
\begin{center}        
  {\Large
 F-Theory GUT Vacua on Compact Calabi-Yau Fourfolds
  }
\end{center}

\vspace{0.75cm}
\begin{center}        
 Thomas W.~Grimm$^{a,}\footnote{grimm@th.physik.uni-bonn.de},$ Sven Krause$^{b,}$\footnote{s.krause@thphys.uni-heidelberg.de}  and Timo Weigand$^{b,}$\footnote{t.weigand@thphys.uni-heidelberg.de}
\end{center}

\vspace{0.15cm}
\begin{center}        
  \emph{$^{a}$ Bethe Center for Theoretical Physics, \\ 
               Nussallee 12, 53115 Bonn, Germany } 
\\[0.15cm]
  \emph{$^{b}$ Institute for Theoretical Physics, University of Heidelberg,  \\ 
               Philosophenweg 19, 69120 Heidelberg, Germany } 
\\[0.15cm]
\end{center}

\vspace{2cm}


\begin{abstract}

We present compact three-generation F-theory GUT models meeting in particular the constraints of D3-tadpole cancellation and D-term supersymmetry. 
To this end we explicitly construct elliptically fibered Calabi-Yau fourfolds 
 as complete intersections in 
a toric ambient space. Toric methods enable us to control the singular geometry of the  SU(5) GUT model.
The GUT brane wraps 
  a non-generic del Pezzo surface  admitting  GUT symmetry breaking via hypercharge flux.
It is contractible to a curve and we demonstrate the
existence of a consistent decoupling limit. We compute the Euler characteristic of the singular Calabi-Yau fourfold 
to show that our three-generation flux solutions obtained via the spectral cover construction are consistent with D3-tadpole cancellation.

\end{abstract}

\clearpage



\newpage
\section{Introduction}




F-theory \cite{Vafa:1996xn} has been known for over a decade to be a fascinating laboratory to study the non-perturbative dynamics of compactifications with branes, but only recently has it  been appreciated also
  as a  corner of the string landscape that is  surprisingly fruitful for string phenomenology \cite{Donagi:2008ca,Beasley:2008dc,Beasley:2008kw,Donagi:2008kj}.\footnote{Recent advances studying various aspects of brane dynamics in F-theory include \cite{Collinucci:2008pf,Braun:2008ua,Braun:2008pz,Grimm:2009ef,Alim:2009bx,Braun:2009bh,GrimmNew,Jockers:2009ti}.}
 The appearance of exceptional gauge groups and  the localisation of the gauge degrees of freedom on branes make it a natural arena for Grand Unified models of particle physics. This is true in particular for the construction of GUT models based on $SU(5)$ dynamics which hold the remarkable  potential to avoid some of the notorious shortcomings   \cite{Murayama:2001ur} that plague purely field theoretic approaches. 
 Among the phenomenological challenges that have been studied  in this framework are the question of GUT symmetry breaking via hypercharge flux and doublet-triplet splitting \cite{Beasley:2008kw,Donagi:2008kj}, the suppression of unacceptable proton decay  \cite{Beasley:2008kw,Hayashi:2009ge}, gauge coupling unification \cite{Donagi:2008kj,Blumenhagen:2008aw,Conlon:2009qa}, aspects of neutrino physics \cite{Tatar:2009jk,Bouchard:2009bu} and even the potential engineering of the observed hierarchical flavour structures \cite{Heckman:2008qa,Randall:2009dw,Font:2009gq,Heckman:2009mn,Cecotti:2009zf,Hayashi:2009bt,Conlon:2009qq,Dudas:2009hu,Marchesano:2009rz}.
 
Some of these topics can actually be successfully addressed in a local setup without specifying the full compactification manifold in detail, but focusing on the neighbourhood of the GUT brane and its geometric properties. Unlike pure bottom-up model building, however, the program of string phenomenology is not restricted to the proposal of promising phenomenological mechanisms. To prove the actual viability of a phenomenological idea it is desirable to study its actual realisation within the landscape of well-defined string vacua. 
 
The explicit construction of compact $SU(5)$ GUT vacua with hypercharge 
induced GUT symmetry breaking \cite{Beasley:2008kw,Donagi:2008kj} was initiated 
in \cite{Blumenhagen:2008zz} actually not in fully fledged F-theory, but, as a first step, 
in the framework of Type IIB orientifolds within the well-defined framework of 
toric geometry. 
Inspired by the fact that the geometries of these 3-generation GUT models allow for a lift to F-theory \cite{Collinucci:2008zs,Collinucci:2009uh,Blumenhagen:2009up}, \emph{explicit} Calabi-Yau fourfolds 
admitting three-generation F-theory GUT models were realised in \cite{Blumenhagen:2009yv}.  
This toric approach to elliptic fourfolds goes beyond the construction of the complex three-dimensional base space. 
As we will review in section \ref{sec_Fourfolds}, a class of base geometries for which the existence of an elliptic fibration over 
this base is guaranteed are certain Fano threefolds. In this case, however, the corresponding F-theory 
model is known to contain only abelian gauge groups.  
However, the engineering of non-abelian $SU(5)$ GUT symmetry in viable models with suitable matter curves
forces the base to be non-Fano \cite{Cordova:2009fg} and
renders the fourfold singular. To keep full control over the geometry, \cite{Blumenhagen:2009yv} directly constructed the resolved Calabi-Yau fourfold with a non-Fano base, thereby putting the existence of the desired $SU(5)$ GUT fourfold on firm grounds. 
Another setup for $SU(5)$ GUT models is presented in \cite{Marsano:2009ym,Marsano:2009gv,Marsano:2009wr}, whose geometric analysis focuses on the construction of a  non-Fano base space.  

The second technical ingredient  of F-theory GUT vacua is non-trivial gauge flux, which is indispensable  in order to achieve a chiral spectrum on the matter curves. Via duality with M-theory  gauge flux in F-theory should be thought of as a special type of $G_4$ background flux with two legs along the 7-brane. New light on the practical implementation of such fluxes was shed in \cite{Donagi:2008ca,Hayashi:2008ba,Tatar:2009jk,Donagi:2009ra} in terms of the spectral cover construction \cite{Friedman:1997yq,donagi-1997-1}. This approach can be understood as a truncation of the Tate model encoding the singular Calabi-Yau fourfold to the neighbourhood of the GUT brane $S$ inside the base $B$. The gauge flux is effectively treated in a manner familiar from heterotic model building on elliptically fibered Calabi-Yau threefolds. While inspired by the treatment of gauge flux in models with a heterotic dual \cite{Curio:1998bva}, there are strong indications that the spectral cover approach is applicable also to more general F-theory models without obvious heterotic duals. In fact, in  \cite{Blumenhagen:2009yv} a simple formula was conjectured, based on the spectral cover construction, for the Euler characteristic of the singular Calabi-Yau fourfolds relevant for F-theory models with non-abelian gauge dynamics. The predictions of this formula could be compared with the values computed directly from the explicit toric resolution of the singular fourfolds, finding perfect match.
Besides lending further credibility to the spectral cover approach as such, this formula also serves as a simple means to calculate the curvature induced D3-brane tadpole contribution of the geometry.\footnote{Alternatively, in \cite{Andreas:1999ng,Andreas:2009uf} the Euler characteristic of singular fourfolds is computed via direct resolution.} We hasten to add, though, that its applicability is guaranteed a priori only for  models with a single non-abelian gauge group along one divisor - a feature which can  be checked at least for explicitly constructed  Calabi-Yau fourfolds as in  \cite{Blumenhagen:2009yv}. 

In this short article, we continue the construction of compact $SU(5)$ GUT models, generalising the approach developed in \cite{Blumenhagen:2009yv}. While that work did construct flux solutions for three-generation models, one of the remaining technical challenges  was the fact that the D3-brane tadpole of the gauge flux exceeded the curvature induced D3-tadpole encoded in the Euler characteristic of the fourfold. This requires the introduction of anti D3-branes. To remedy this problem, we generalise the geometries of \cite{Blumenhagen:2009yv} to an entire sequence of torically constructed Calabi-Yau fourfolds. 
This leads to the construction of a number of compact three-generation $SU(5)$ GUT models which allow for the consistent cancellation of 3-brane tadpoles without sacrificing the explicit control of the geometry of the fourfold.
The models are realised in terms of a \emph{split} spectral cover, which was shown in \cite{Tatar:2009jk} to be the appropriate framework for $SU(5)$ GUT models to avoid problems with proton decay. The consequences of this split were analysed in detail in \cite{Marsano:2009gv}. As in \cite{Blumenhagen:2009yv} we describe the gauge flux of the split spectral cover as an $S[U(4)\times U(1)]$ bundle, paying particular attention to a correct quantisation and the D-term supersymmetry conditions. 

Given the nature of the GUT divisor as the blow-up of curves of del Pezzo singularities, it is contractible not to a point, but to a curve inside the base $B$ of the elliptic fibration, at least if we concentrate on situations where the volume of $B$ remains finite.
We explicitly demonstrate the physical viability of such a situation by examining the possibility of decoupling  gravity from the GUT theory on the brane. Such a decoupling limit goes beyond the minimal requirements for the realisation of the  little hierarchy between the GUT and the Planck scale, but has been argued to be a reasonable organising principle for GUT model building with branes \cite{Donagi:2008ca,Beasley:2008dc,Cordova:2009fg}. Our analysis shows that if the GUT brane is contractible to a curve in a limit that keeps the volume of the ambient space finite, gravity can in general be  consistently decoupled without leaving the realm of the supergravity approximation. 

This note is organised as follows:
In section \ref{sec_Fourfolds} we describe our toric approach to the construction of elliptically fibered Calabi-Yau fourfolds with a contractible divisor as candidates for $SU(5)$ GUT models in F-theory, leaving some of the technical details to the appendix \ref{appToric}.
In particular, in subsection \ref{sec_GeomEx} we  specialise to one particular example based on a fibration over $\mathbb P^4[3]$ on which we build a three-generation GUT model. The details of its toric data can be found in appendix \ref{appExample}. This fourfold is merely one example in a whole class of geometries, and a related geometry is presented in appendix \ref{appExample2}.
Section \ref{Decoupling} demonstrates the possibility of taking a consistent  decoupling limit for gravity by shrinking the GUT brane to a curve inside the fourfold. We then review the main features of the spectral cover construction in section \ref{sec_SCC}. This puts us in a position to present, in section \ref{sec_3Gen}, a three-generation GUT model that consistently cancels all tadpoles, and is D-term supersymmetric inside the K\"ahler cone in a manner consistent with the decoupling limit. In appendix \ref{appExample} we provide several more such three-generation flux models, while some comparable solutions for the cousin geometry are collected  in appendix \ref{appExample2}.

\section{Construction of Calabi-Yau fourfolds}\label{sec_Fourfolds}

In this section we describe the explicit construction 
of compact elliptically fibered Calabi-Yau fourfolds $Y$ 
which we will use for GUT model building in the later sections. 
We will denote the base of this fibration by $B$.
Recall that F-theory on $Y$ provides a geometrisation of 
an $\cN = 1$ Type IIB compactification on $B$ 
with holomorphically varying complexified string coupling 
constant $\tau = C_0 + i e^{-\phi}$ \cite{Vafa:1996xn}. 
The parameter $\tau$ is identified with 
the complex structure modulus of a two-torus and varies over 
the K\"ahler manifold $B$.
This complex fourfold geometry captures non-trivial
monodromies of $\tau$ around degeneration loci of the two-torus. Precisely this provides
a geometrisation of seven-branes. Note that this degeneration 
can be so severe that not only the elliptic fibration becomes singular, but 
also the fourfold $Y$ itself. This indicates the presence of a non-abelian
gauge group on the 7-branes. Since we are interested in GUT model building we 
will need to degenerate the fibration to obtain an $SU(5)$ gauge group.
To nevertheless work with a smooth space, one can paste in resolving $\bbP^1$ fibers.
It will be this resolved space which we will construct explicitly in this section. 
The F-theory limit states that one eventually has to take the volume of 
the elliptic fiber and the resolving $\bbP^1$'s to zero. 

The existence of a compact elliptically fibered Calabi-Yau fourfold implies that 
the 7-brane tadpoles are automatically cancelled. This is not the case for the 
D3-brane tadpoles. A globally consistent setup has to satisfy \cite{Sethi:1996es}
\beq \label{chi/24}
   \frac{\chi(Y)}{24} = n_{\rm D3} + \frac{1}{2} \int_{Y} G_4 \wedge G_4\ ,
\eeq
where $\chi(Y)$ is the Euler number of $Y$, $n_{\rm D3}$ is the number of movable
D3-branes and $G_4$ is the four-form flux. In order to avoid uncontrolled  supersymmetry breaking
it is desirable to only include D3-branes into the set-up such that $n_{\rm D3} \ge 0$. 
Hence, at least from these simple considerations of tadpole cancellation, it appears 
desirable to study Calabi-Yau fourfolds with a large $\chi(Y)$.

\subsection{Calabi-Yau fourfolds with abelian gauge symmetries}
\label{sec_Ab}

To explicitly construct an elliptically fibered $Y$ one might be tempted to simply start with a 
base $B$ and specify $Y$  via the Weierstrass equation 
\beq \label{Weierstrass}
   P_{\rm W}(x,y,z|y_i) = y^2 - x^3 + f z^4 + g z^6 = 0\ 
\eeq
or, equivalently, the Tate form 
\beq \label{Tate1}
    P_{\rm W} = x^3 - y^2 + x\, y\,  z\, a_1 + x^2\, z^2\, a_2 + y\, z^3\,a_3   + x\, z^4\, a_4 + \, z^6\, a_6\ = 0.
\eeq
Here $f,g,a_i$ depend on the  coordinates $y_i$ of the base $B$, while the generic elliptic fiber $\bbP_{1,2,3}[6]$ has projective coordinates $(z,x,y)$. This construction 
can be carried out for base spaces $B$ over which there exists an elliptic 
fibration with no non-abelian singularities. However, to build
an $SU(5)$ GUT model such a construction will be much more involved and one has to 
face the challenge of constructing $Y$ directly including its resolved singular fibers. 

Before turning to the explicit construction of $Y$ with non-abelian singularities, 
we recall a few facts on base spaces which do admit elliptic fibrations  not rendering $Y$ itself singular.
Such $B$ can be found among the
finite list of Fano threefolds \cite{MoriMukai1,MoriMukai2}. Let us recall four examples 
with a single K\"ahler class:  
\beq \label{small_list}
  \begin{array}{c|cccc} & \quad \bbP^3 \quad & \quad \bbP^4[2]  \quad & \quad\bbP^4[3] \quad & \quad \bbP^4[4]\\[.1cm] \hline
                        \rule[0cm]{0cm}{.5cm} c^3_1(B) & 64 & 54 & 24 & 4 \\
                        \rule[0cm]{0cm}{.5cm} h^{2,1}(B) & 0 & 0 & 5 & 30 \\[.1cm] \hline
                        \rule[0cm]{0cm}{.5cm} \chi(Y) & 23328 & 19728 & 8928 & 1728 \\                       
  \end{array}
\eeq
Here $\bbP^4[n] = \{P_n(y_1,\ldots,y_5) = 0\}$ denote generic hypersurfaces 
of degree $n$ in the projective space $\bbP^4$. The $\bbP^4[n]$ in \eqref{small_list}
are the quadric, cubic and quartic Fano threefolds. The famous quintic threefold would be 
the next in the list but is Calabi-Yau and thus does not admit an elliptic fibration.
Note that in \eqref{small_list} we have also displayed the Euler numbers of the 
associated fourfolds $Y$ obtained by an elliptic fibration evaluating \cite{Sethi:1996es,Klemm:1996ts}
\beq \label{smooth_chi}
    \chi(Y) = 12 \int_B c_1(B) c_2(B) + 360 \int_B c_1^3(B)\ .
\eeq  
Due to the special properties 
of the base spaces $B$ listed in \eqref{small_list} 
the elliptic fibrations $Y$ are known to exist.\footnote{Here one uses the fact 
that these $B$ admit a very ample canonical bundle. This is not true for all 
Fano threefolds, see ref.~\cite{Grassi:1997mr}.} In fact, we will show in the 
next section that the fibration $Y$ can be constructed explicitly.

From the so constructed fourfolds one can proceed by performing geometric 
transitions. More precisely, one can fix some of the $h^{3,1}(Y)$ complex structure deformations 
 to generate a singularity in $B$ and blow this up into a divisor. 
From the perspective of GUT model building it is natural to blow up singular curves
in $B$ \cite{Donagi:2009ra,Blumenhagen:2009yv,Cordova:2009fg}. In this blow-up process one replaces the 
singular curve with an exceptional divisor $E$ thus yielding a new base $\tilde B$.
The canonical class of the base changes under this transformation as
\beq
   K_{\tilde B} = K_{B} + E\ .
\eeq  
Using this blow-up procedure we will generate del Pezzo surfaces in the base $\tilde B$.
Recall that del Pezzo surfaces are by definition precisely the two-dimensional Fano manifolds. 
The $10$ del Pezzo surfaces are $\mathbb{P}^1 \times \mathbb{P}^1$ and $dP_n$, 
which is $\mathbb{P}^2$ with $n=0,\ldots,8$ points blown up to $\mathbb{P}^1$'s.
The non-toric del Pezzos are realised 
as complete intersections or hypersurfaces in a projective or toric 
ambient space, just as the Calabi-Yau fourfolds themselves. For example
one can represent 
\beq \label{dP678_hyper}
 dP_5 = \bbP_{11111}[2,2],\qquad dP_6 = \mathbb{P}_{1111}[3] , \qquad dP_7 = \mathbb{P}_{1112}[4] , \qquad dP_8 = \mathbb{P}_{1123}[6] ,
\eeq
where the subscripts denote the weight of the $4$ or $5$ projective coordinates and we have also indicated the 
degree of the hypersurfaces in the square brackets. Note that $dP_5$ is the complete intersection of two quadratic  
constraints in $\bbP^4$. After the del Pezzo transition, the base $\tilde B$
will generically no longer be Fano and the existence of an elliptically fibered $Y$ has to be shown.\footnote{For ease of notation we will henceforth stick to the symbol $B$ also for the non-Fano base.} 
Therefore, as in ref.~\cite{Blumenhagen:2009yv}, we will construct $Y$ directly in section \ref{explicit_4folds}.

\subsection{Calabi-Yau fourfolds in toric ambient spaces}
\label{explicit_4folds}

Let us stress that the procedure of section \ref{sec_Ab} to construct a Calabi-Yau fourfold simply as the Weierstrass model over a base $B$ will in general not be sufficient
for every $B$.  The reason is that the geometry of $B$ will often enforce a non-trivial degeneration of the elliptic fibration. E.g. the computation of the relevant Euler characteristic, either directly \cite{Andreas:1999ng,Andreas:2009uf} or via the method proposed for a certain class of geometries in \cite{Blumenhagen:2009yv}, requires explicit control of the singularities of $Y$. Hence, if we want to construct a compact F-theory GUT model 
we will need to define the fourfold $Y$ directly.  The fact that
$Y$ and $B$ are complex K\"ahler manifolds provides us with the necessary 
powerful tools to construct and study these manifolds as we will recall in the sequel. 

The best studied technique to explicitly construct Calabi-Yau manifolds $Y$ 
is to realise these as subvarieties in a higher-dimensional 
ambient space. In the sequel these ambient spaces are 
either projective spaces or toric spaces which we denote by $\bbP_{\Delta}$.
The Calabi-Yau spaces $Y$ are specified by a number of complex equations inside
$\bbP_{\Delta}$. As projective spaces, toric spaces can be described by 
a set of coordinates $x_1,\ldots,x_n$ modulo a number of 
scaling relations $x_i\mapsto (\mu_{k})^{\ell^{k}_{i}}\ x_i$ with $\mu_{k} \in \bbC^*$. 
In the language of the linear sigma model one can hence view the $x_i$ as fields which are charged under a number $r$ of 
complexified $U(1)_k$ gauge groups with charges $\ell^{k}_{i},\ k=1,...,r$.
After we discard disallowed points $Z$, in analogy to the origin in a projective space,
modding out  the complex gauge group actions specified by $\ell^k$ 
defines the toric ambient space
\beq
\label{PDelta}
  \bbP_{\Delta} = (C(x_1,\ldots,x_n) - Z)/(\bbC^*)^r\ 
\eeq 
of dimension dim($\bbP_{\Delta}$)$=n-r$.
The divisors $D_i$ of $\bbP_{\Delta}$ are 
given by the equations $x_i=0$, which allows us to write the first Chern class 
of $\bbP_{\Delta}$ as $c_1(\bbP_{\Delta}) = \sum_i D_i$.
The  manifold $Y$ itself is 
given by complex polynomial equations, $p_a(x_1,\ldots,x_n) = 0$, 
which transform consistently under the scalings
$\ell^k$. In order for $Y$ to be Calabi-Yau, i.e.~$c_1(Y)=0$, 
the class of the product of the $p_a$ has to be $c_1(\bbP_{\Delta})$, such that 
\beq \label{nef-partition}
    c_1(\bbP_{\Delta}) = \underbrace{(D_1 + \ldots + D_{a_1})}_{[p_1]} + \underbrace{(D_{a_1 + 1}+ \ldots + D_{a_2})}_{[p_2]} + \ldots\ ,
\eeq
where $[p_i]$ is the class of the polynomial $p_i$.
Hence, together with a consistent specification of the partitions of $c_1(\bbP_{\Delta})$ 
corresponding to the polynomials $p_i$, the ambient space $\bbP_{\Delta}$ canonically 
specifies a Calabi-Yau space $Y$ in $\bbP_{\Delta}$. 
The specification of a consistent partition \eqref{nef-partition} makes 
the construction of complete intersections in a toric ambient space more involved 
compared to the case of hypersurfaces \cite{Borisov-1993, batyrev-1994-1, batyrev-1994-2, Kreuzer:2001fu, Klemm:2004km}. 
This can be traced back to the fact that the intersection has to be transversal 
in order that $Y$ be non-singular.

In appendix \ref{appToric} we describe the construction of such complete intersections in terms of reflexive polyhedra.
The analysis of these data yields the necessary tools to compute the topological 
information of the Calabi-Yau fourfold. In the rest of this article 
we will only concentrate on  elliptically fibered $Y$ given by two 
equations
\beq
   P_{\rm W}(x,y,z|y_i) = 0 \ , \qquad P_B(y_i)=0\ ,
\eeq
where we have now specified the coordinates $(x,y,z)$ of the elliptic fiber 
among the $x_i$. The Calabi-Yau condition demands that the class of $P_{\rm W} \cdot P_B$
is $c_1(\bbP_{\Delta})$.
Note that without an additional $P_B$ the manifold 
$Y$ is determined by one constraint and is a hypersurface. 
There exists a large set of elliptically 
fibered Calabi-Yau fourfold hypersurfaces. However, as we argued in ref.~\cite{Blumenhagen:2009yv}
and we will see again below, the inclusion of $P_B$ provides us with 
a richer set of geometries for the GUT seven-branes. In particular, 
in this way we can place the GUT brane 
on non-toric del Pezzo surfaces.

Simple complete intersections can be constructed by considering an 
elliptic $\bbP_{123}[6]$ fibration over the Fano threefolds \eqref{small_list}. 
In appendix \ref{appToric} we give the details of this construction for 
the Calabi-Yau threefold with Fano threefold base $\mathbb P^4[3]$. 
An example originating from the Fano threefold $\bbP^4[4]$ in 
\eqref{small_list} was already constructed in ref.~\cite{Blumenhagen:2009yv}.
We can now proceed to systematically generate singular curves in $B$ and 
blow them up into del Pezzo surfaces. This process will lower the number of 
complex structure moduli of $Y$ and yield new elements of $H^{2}(Y)$ upon 
resolution of the singular curves. 
Again, these transitions are best discussed in the framework of toric 
geometry as we briefly recall in appendix \ref{appToric}. More details on this construction can 
also be found, for example, in refs.~\cite{Blumenhagen:2009yv,GrimmNew}. 
Crucially, we note that the base threefolds obtained after the transition 
will generically be no longer Fano. The existence of an elliptic fibration 
over these new base spaces is, however, guaranteed in the explicit toric 
constructions. One can evaluate the relevant topological 
data such as intersection numbers and Chern classes by applying the 
methods described in refs.~\cite{Borisov-1993, batyrev-1994-1, batyrev-1994-2, Kreuzer:2001fu, Klemm:2004km}.\footnote{%
Here the programs PALP \cite{Kreuzer:2002uu}, TOPCOM \cite{Rambau:TOPCOM-ICMS:2002} and Schubert make these calculations tractable.}

Having constructed Calabi-Yau fourfolds with a GUT del Pezzo surface in the base $B$ 
we have to ensure that the elliptic fiber degenerates appropriately to 
account for the non-abelian GUT gauge group. Since we are interested in a model with $SU(5)$ degeneracy, we have to specify the polynomials $a_n$ appearing in the Tate form \eqref{Tate1} further. If the del Pezzo surface is 
given by the constraint $w=0$ in $B$ one has to realise $a_n$'s of the form 
\beq \label{TateSU5}
  a_1 = \mathfrak{b}_5  , \quad
  a_2 = \mathfrak{b}_4 w , \quad
  a_3 = \mathfrak{b}_3 w^{2} , \quad
  a_4 =  \mathfrak{b}_2 w^{3} , \quad
  a_6 =  \mathfrak{b}_0 w^{5} ,
\eeq
where the functions $\mathfrak{b}_n(w,y_i)$ do not factorise further in the 
$w$. The form \eqref{TateSU5} can be realised by systematically dropping 
monomials from the original Tate constraint. This is readily implemented 
torically by manipulating reflexive polyhedra as we describe in appendix \ref{appToric}. 
The toric calculus automatically resolves the singular fibers. For the 
example considered in the next section, we are thus able to explicitly 
construct a Calabi-Yau fourfold $Y^{SU(5)}$ with resolved $SU(5)$ singularity 
over the GUT brane. Since $Y^{SU(5)}$ is a smooth fourfold we can again compute 
the intersection numbers and Chern classes. The crucial point in the 
model building will be that the values of $\chi(Y)$ of the two Calabi-Yau fourfolds with 
base $\bbP^4[3]$ and $\bbP^{4}[4]$ differ by a factor of $5$ as seen in \eqref{small_list}. 
This value will still drop for $Y^{SU(5)}$. However, as we will show in the next sections, 
models based on transitions of $\bbP^4[3]$ will allow us to satisfy the tadpole constraint 
\eqref{chi/24} without introducing anti-D3-branes.

\subsection{Example: A model based on ${\mathbb P}^4[3]$} \label{sec_GeomEx}

Let us focus now on a concrete compact Calabi-Yau fourfold out of the larger class of models described in the previous section.  
Concretely we consider the blow-up of ${\mathbb P}^4[3]$ over singular curves to produce a $dP_5$ surface. 
The complete set of toric data of this model is presented in appendix \ref{appExample}. More precisely, 
it will be shown there that one can construct the Calabi-Yau fourfold $Y$ in terms of a reflexive 
polyhedron with two nef-partitions (see \Tableref{exampleData}).
The characteristic topological data of this fourfold are
\beq
  \chi(Y) = 6768 \ , \qquad h^{1,1}(Y)=5\ , \qquad h^{2,1}(Y)=0 \ , \qquad h^{3,1}(Y)=1115\ .
\eeq
Using toric methods described in appendix \ref{appToric} one can explicitly resolve an 
$SU(5)$ singularity over the $dP_5$ surface in the base $B$. One thus constructs the 
manifold $Y^{SU(5)}$ and computes its topological data. In particular, one computes 
$\chi(Y^{SU(5)})=5718$.

Let us next have a closer look at the base $B$ of $Y$. 
As every toric hypersurface, the base $B$ admits an interpretation in terms of the linear sigma model.
The corresponding $U(1)$ charges of the toric  divisors are given in table \ref{MoriA}.
\bea 
  \label{MoriA} 
  \renewcommand{\arraystretch}{1.1} 
  \renewcommand{\arraycolsep}{7pt} 
  \begin{array}{|c|c|c|c|c|c|c|c|c||c|} 
  \hline 
           &y_1 & y_2 & y_3 & y_4 & y_5 & y_6 & y_7 &  y_8 & P_B \\ \hline\hline 
  S      &0 & 0 & 1 & 1 & 0 & 0 & -1 & 0 & 0 \\ 
  Y_1 &1 & 0 & 0 & 1 & 1 &  1 & -1& 0 & 2 \\ 
  Y_2 &0 & 1 & 0 & 0 & 0 & 0 & 1 &  0 &   1\\ 
  Y_3 &0 & 0 & 0 & 0 & 1 & 0 &- 1 &  1 &  1\\ 
  \hline 
  \end{array} 
\eea 
Here $\{S, Y_1,Y_2,Y_3\}$ denote the independent divisor classes as elements of $H_4(B, \mathbb Z)$. For example, the divisor $y_7=0$ lies in the class $[y_7=0]= -S - Y_1 + Y_2 - Y_3$. The last column denotes the class of the constraint equation $P_B(y_i)$ that determines $B$ as a hypersurface  $P_B(y_i)=0$ in the ambient toric space.
From this one obtains $c_1(B) = (S + Y_1 + Y_2)$ and since $c_1(N_S) = S$, one can use $c_1(B) = c_1(S) + c_1(N_S)$ to find $c_1(S) = Y_1 + Y_2$. The $dP_5$ surface obtained as the blow-up of a singular curve is given by $[y_3=0]$. 
 The fact that this indeed defines a non-generic $dP_5$ is demonstrated in appendix \ref{appExample} by restricting  the hypersurface polynomial $P_B$ to this surface.
 The non-zero intersection numbers for the independent divisor subset are
\beq
\label{intform}
  \bal
    &S^2 Y_2 = -2,\quad &&S^2 Y_3 = -1,\quad && S Y_1 Y_2 = 2,&&\quad S Y_1 Y_3 = 1,&&\quad S Y_3^2 = -1\\
    &Y_1Y_2^2 = 3, \quad &&Y_1Y_2Y_3 = 1, \quad &&Y_2^3 = 3, &&\quad Y_2Y_3^2 = -1.
  \eal
\eeq\\
For later purposes we also need the K\"ahler cone. A possible basis is given by 
\bea
\label{Kcone}
K_1 = Y_1 + Y_2,  \quad K_2 = Y_2, \quad K_3 = S + Y_1, \quad K_4 = Y_1 + Y_3.
\eea
 The corresponding intersection numbers are
displayed in appendix \ref{appExample}. Indeed one can check that an expansion of the K\"ahler form $J = \sum r_i K_i$ with $r_i > 0$ leads to positive volume for every effective curve on $B$. On the boundary of the K\"ahler cone some $r_i = 0$, signalling the vanishing of the volume of a curve.

The GUT brane of our model will be placed on the $dP_5$ divisor $y_3=0$ in the class $S$.
To study the embedding of this divisor into the ambient geometry we need to analyse the relation between the  (co)homologies on $B$ and on $S$.  This is important e.g. because GUT symmetry breaking will be implemented via hypercharge flux ${\cal L}_Y$ through 2-cycles on $S$ which are homologically trivial, i.e. the boundaries of 3-chains, on $B$.

The homology group $H_2(S, \mathbb Z )$ of the $dP_5$ surface $S$ is spanned by the hyperplane class $l$ together with the classes $E_i, i=1, \ldots 5$, of the five blow-up ${\mathbb P}^1$s inside $dP_5$.
In terms of these, 
\beq \label{eqc1S}
c_1(S) = 3 l - \sum_{i=1}^5 E_i .
\eeq
The embedding $\iota$ of $S$ into $B$ induces a map $\iota^{!}: H_4(B) \rightarrow H_2(S)$, which assigns to every divisor $D$ on $B$ a curve $\iota^{!} D = D|_S$ on $S$. This curve is given by the intersection of the divisor with $S$, i.e. the restriction of the divisor to $S$.
From the intersection form (\ref{intform}) the pullback map ${\iota}^!: H_4(B, \mathbb Z) \rightarrow H_2(S, \mathbb Z)$ is only of rank 3. This is because one of the four divisor classes on $B$, namely $S+Y_1$, does not intersect $S$ in a non-trivial class. 
More precisely,  $S (S+Y_1) A = 0$ for all divisors $A \in H_4(B, \mathbb Z)$, so that there are only three nontrivial, linearly independent divisors intersecting $S$.\\
It follows that only three elements in $H_2(S, \mathbb Z)$ are non-trivial also as classes of $H_2(B,\mathbb Z)$.
For the remaining elements trivial in the ambient homology, we require that they are perpendicular to $c_1(S)$
as well as to the non-trivial curves, which will later on be determined appropriately. Without loss we choose the basis of trivial elements to be $E_2 - E_3$, $E_3 - E_4$, and $E_4 - E_5$. Then clearly the coefficients of $E_i$ are the same for $i \in \{2,..,5\}$ for all non-trivial elements. Defining $E_I = \sum_{i=2}^5 E_i$, and deducing the form of the non-trivial base elements from the intersection numbers, we find, up to isormorphisms of $dP_5$,
\beq
\label{pullback1}
  \bal
    C_1 &= (Y_1 - S)|_S = 2 l - 2 E_1,\\
    C_2 &= Y_2|_S = 2 l - E_I,\\
    C_3 &= Y_3|_S = 2 l - E_1 - E_I.
  \eal
\eeq
Combining this information with \eqref{eqc1S} then leads to
\beq
\label{pullback2}
S|_S = -l + E_1, \qquad Y_1|_S= l - E_1.
\eeq

\section{Decoupling Limit} \label{Decoupling}

In this section we discuss and clarify certain aspects of the decoupling limit of F-theory GUT models on geometries of the type presented above.
Assume therefore that the divisor $S$ carries a grand unified $SU(5)$ gauge symmetry.
A physically acceptable model must account for the little hierarchy between the Planck scale $M_{Pl} = 1.2 \times 10^{19} \, {\rm GeV}$ and the GUT scale $M_{\rm GUT} = 3 \times 10^{16} \, {\rm GeV}$.
These scales are related to the volumes of the GUT divisor $S$ and the threefold $B$ via
\bea
M_{Pl}^2 = M_{\ast}^8 \, {\rm Vol}(B), \quad\quad M^4_{\rm GUT} \simeq  {\rm Vol}^{-1}(S).
\eea
The first assertion follows from dimensional reduction of the ten-dimensional IIB action in Einstein frame with inverse string scale $M_{\ast} = \ell_s^{-1}$; the second estimate reflects the breaking of the GUT gauge group by hypercharge flux \cite{Beasley:2008kw,Donagi:2008kj}, which is quantised in units of the volume of the 2-cycles on $S$, and is thus true for an approximately isotropic divisor $S$.
In addition, the gauge coupling $\alpha_{GUT}$ at the GUT scale is, to first order, given by
\bea
\frac{1}{\alpha_{GUT}} = M^4_{\ast}\,  {\rm Vol}(S) \simeq 24.
\eea
The typical length scales $R_S$, $R_B$ defined by  $R_S^4 ={\rm Vol}(S)$ and $R^6_{B} = {\rm Vol}(B)$ are thus subject to the comparatively mild tuning of roughly
\bea
\label{tuning}
 R_S \simeq 2.2 \, \ell_s, \quad\quad R_B \simeq 5.6 \, \ell_s, \quad\quad \ell_s \simeq  0.15 \times 10^{-16}\,  {\rm GeV}^{-1} .
\eea
The  minimal requirement for a physically acceptable GUT vacuum is therefore that a K\"ahler form $J$ inside the K\"ahler cone can be found which leads to the relation (\ref{tuning}) between ${\rm Vol}(B)$ and ${\rm Vol}(S)$. In addition, this relation must be achieved without pushing any cycle volume to the string scale since in this case $\alpha'$ corrections would not be under control any longer. While eventually a mechanism must be found that stabilises the K\"ahler moduli dynamically in that regime, a first consistency condition on the geometry is whether such a parameter regime inside the K\"ahler cone exists at all.\footnote{The presence of gauge fluxes generically leads to extra constraints that must be satisfied in this regime such as D-term supersymmetry conditions.}

A much stronger requirement is that the Planck scale can be taken to infinity while keeping $M_{\rm GUT}$ fixed. This means that gravity can be consistently decoupled from the gauge theory of the GUT brane altogether, as is motivated by the apparent UV completeness of the GUT theory. 
While, strictly speaking, it would suffice for a physically consistent GUT model to satisfy the much milder constraints (\ref{tuning}), the existence of a decoupling limit has been emphasised in the recent literature \cite{Verlinde:2005jr,Beasley:2008kw} as an organising principle of local (GUT) model building. 

This strict decoupling limit $M_{Pl} \rightarrow \infty$ while $M_{GUT}$ finite amounts to finding a limit in which ${\rm Vol}(B) = J^3 \rightarrow \infty$ while keeping at the same time ${\rm Vol}(S) = S J^2$ constant. Alternatively one can consider the limit  ${\rm Vol}(S) = S J^2 \rightarrow 0$,  ${\rm Vol}(B) = J^3$ constant, even though we  will exemplify momentarily that some care is to be taken in  equating the two pictures. These two different viewpoints will be referred to as physical versus mathematical decoupling.\footnote{~Our definition of the mathematical decoupling limit differs from the one considered in \cite{Cordova:2009fg}: We insist that ${\rm Vol}(B) = J^3$ and therefore $M_{Pl}$ remains finite in the limit in which the volume of $S$ shrinks to zero.}

At first sight, the simplest possibility of the mathematical decoupling limit would correspond to a GUT brane that can shrink to a \emph{point} inside a finite threefold $B$. This would be the case if $S$ wraps a completely generic Fano surface inside a non-Fano base $B$ \cite{Cordova:2009fg}. In the compact examples of the type discussed  in \cite{Blumenhagen:2009yv}  and in this article, by contrast, $S$ is a non-generic del Pezzo surface and is only contractible to 
 a \emph{curve}. This will be the matter curve on which the ${\bf 10}$ representation is localised. More generally, \cite{Cordova:2009fg} argued that the consistency conditions of an SU(5) GUT model with chiral matter are  compatible only with shrinkability to a curve or a non-generic situation in which contractibility of the GUT brane implies also that some other divisor intersecting $S$ in a curve must shrink. 
 
 One might wonder  whether the first option, contractibility to a curve while $\rm{Vol}(B)$ finite, leads to a physically acceptable GUT model. A potential worry  might be that 
 the process ${\rm Vol}(B) \rightarrow \infty$ with ${\rm Vol}(S)$ finite might also take the volume of some curve inside $S$ to infinity. To keep the volume of $S$ finite, this would mean that
 some direction normal to this  curve inside $S$ is necessarily pushed to the stringy or sub-stringy regime where the supergravity approximation breaks down. In the remainder of this section we show explicitly that this worry is in general not justified, thereby demonstrating  the physical viability of our geometries for GUT model building.

\subsubsection*{Contractibility to a curve is physically acceptable}

To analyse the decoupling properties of the GUT geometry introduced in section \ref{sec_GeomEx}
we start by expanding the K\"ahler form $J$ into the generators of the K\"ahler cone, $J = \sum_{i=1}^4 r_i K_i$, $r_i > 0$. From the intersection numbers for the K\"ahler cone \eqref{KaehlerIntersec}  one finds that
\beq
  \bal
    {\rm Vol}(B) = J^3 = &\ r_1^2 (12 r_1 + 9 r_2 + 7 r_3 + 5 r_4)\\
    			& + r_1 (6 r_2^2 + 5 r_2 r_3 + 4 r_2 r_4 + 2 r_3^2 + 4 r_3 r_4 + r_4^2)\\
			& + r_2^2 (3 r_2 + 3 r_3 + 3 r_4) + r_2 ( 2 r_3^2 + 3 r_2 r_3 + r_4^2)\\
			& + r_3^2 r_4 + r_3 r_4^2
  \eal
\eeq
and
\beq
{\rm Vol}(S) = S J^2 = r_1 (4 r_1 + 4 r_2 + 6 r_4) + r_4 (4 r_2 + r_4) .
\eeq
Let us consider the mathematical and physical decoupling limit in turn.
In the first, the volume of the GUT divisor $S$  
vanishes for $r_1 \rightarrow 0, r_4 \rightarrow 0$ while the volume of $B$
remains finite as long as $r_2$ remains finite. Note that $r_3$ is unconstrained by these two requirements.
It is easy to see that $S$ can only shrink to a curve of finite volume.
In fact, consider the curve $C_1  = S \cap Y_1$. Its volume  $S \,  Y_1 \, J =2 r_1 + 2 r_2 + r_4 $  cannot vanish if ${\rm Vol}(B)$ and thus $r_2$ is to stay finite.

On the other hand, to determine the phenomenology of the decoupling limit, one should actually consider the inverse, physical limit in the sense defined above.
Now the volume of $S$ and thus $M_{GUT}$ remains finite for finite $r_1$, $r_2$ and $r_4$ and ${\rm Vol}(B) \rightarrow \infty$ requires $r_3 \rightarrow \infty$.
This demonstrates explicitly that in general the mathematical and the physical decoupling picture are controlled by different vectors in the K\"ahler cone. 
 In particular, the fact that $S$ can only shrink to a curve (in the mathematical picture) does by no means imply the existence of a curve inside $S$ whose volume would diverge in the physical picture ${\rm Vol}(B) \rightarrow \infty$ while ${\rm Vol}(S)$ finite.\footnote{Note that our conclusions are not in contradiction to those in \cite{Cordova:2009fg}, which considered a different type of mathematical decoupling limit.} Indeed the existence of such a curve would imply that in the physical decoupling limit the neighbourhood normal to this curve inside $S$ would have to acquire sub-string length, questioning the applicability of perturbation theory for the GUT theory on $S$. Our analysis shows, however, that in general divisors shrinkable to a curve are perfectly fine for GUT model building as this phenomenon need not occur. In fact one immediately checks that no curve inside $S$ exists whose volume is controlled by $r_3$ because the pullback of the associated K\"ahler class $K_3$ to $S$ vanishes.
At the same time we see that $M_{GUT}$ is dependent not just on a single K\"ahler modulus, but in this case on three K\"ahler moduli $r_1, r_2, r_4$, all of which have to be fixed at a scale not exceeding the GUT scale. This is different from situations where the volume of the GUT brane is controlled by a single modulus that sets the scale of $M_{GUT}$ and might make a dynamical stabilisation of the GUT scale harder.

 In order to determine whether  any other divisors stay at finite volume in the physical decoupling process, consider the volumes of the other linear independent divisors, 
\beq
  \bal
    {\rm Vol}(Y_1) = Y_1 J^2 = &\ r_1 (3 r_1 + 6 r_2 + 4 r_3 + 2 r_4) + r_2 (3 r_2 + 4 r_3 + 2 r_4) + 2 r_3 r_4, \\
    {\rm Vol}(Y_2) = Y_2 J^2 = &\ r_1 (9 r_1 + 12 r_2 + 10 r_3 + 8 r_4)\\
    &+ r_2 (3 r_2 + 6 r_3 + 6 r_4) + r_3 (2 r_3 + 6 r_4) + r_4^2,\\
    {\rm Vol}(Y_3) = Y_3 J^2 = &\ r_1 (2 r_1 + 2 r_2 + 4 r_3) + 2 r_2 r_3 + r_3^2 .
  \eal
\eeq
By inspection of the various coefficients of the $r_3 r_i$-terms one finds that there is no linear combination of $Y_i$ whose volume is independent of $r_3$, so that sending the latter to infinity only leaves the divisor $S$ at finite volume.

\section{Compact Calabi-Yau fourfolds and spectral covers} \label{sec_SCC}

To make this paper as self-contained as possible we use this section to recall the necessary ingredients for the construction of an F-theory GUT model based on the previously introduced Calabi-Yau fourfolds. This review closely follows the presentation in  \cite{Blumenhagen:2009yv} and we refer to \cite{Donagi:2008ca,Hayashi:2008ba,Hayashi:2009ge,Donagi:2009ra,Marsano:2009gv}  for further details of the spectral cover in this specific context.

\subsection{Complete-intersecting fourfolds and the Tate form} \label{Tateform}

Recall that we explicitly realise the Calabi-Yau fourfold $Y$ 
via \textit{two} hypersurface constraints 
\beq \label{two_constr}
   P_B(y_i) = 0 ,\qquad P_{\rm W}(x,y,z;y_i)=0 
\eeq
in a six-dimensional projective or toric ambient space. Here $P_B$
is the constraint of the base $B$ which is independent of the coordinates 
$(x,y,z)$ of the elliptic fiber.\footnote{Note that this more general setting 
also includes hypersurfaces encoded by a single constraint $P_{\rm W} =0$ if 
we choose $P_B$ to be trivial. } 
The Weierstrass polynomial $P_{\rm W}$  encodes the structure of the elliptic fibration and can be brought into the Tate form 
\beq \label{Tate1}
    P_{\rm W} = x^3 - y^2 + x\, y\,  z\, a_1 + x^2\, z^2\, a_2 + y\, z^3\,a_3   + x\, z^4\, a_4 + \, z^6\, a_6\ = 0,
\eeq
where $(x,y,z)$ are coordinates of the torus fiber. In the sequel we will only 
be working with the inhomogeneous Tate form by setting $z=1$. The $a_n(y_i)$ are 
sections of $K_B^{-n}$, with $K_B$ being the canonical bundle of the base $B$.  
If one sets all $a_n=1$ equation  \eqref{Tate1} reduces to the elliptic fiber $\bbP_{123}[6]$. 

The $a_n$ encode the discriminant of the elliptic fibration. In terms of the new sections 
\beq
  \beta_2 = a_1^2 + 4 a_2 ,\qquad 
  \beta_4 = a_1 a_3 + 2\, a_4 ,\qquad
  \beta_6 = a_3^2 + 4 a_6  ,
\eeq
the discriminant can be expressed as
\beq
  \Delta = -\tfrac14 \beta_2^2 (\beta_2 \beta_6 - \beta_4^2) - 8 \beta_4^3 - 27 \beta_6^2 + 9 \beta_2 \beta_4 \beta_6 ,
\eeq
which is a section of $K_B^{-12}$. In general, the discriminant $\Delta$ will 
factorise with each factor describing the location of a 7-brane on a divisor 
$D_i$ in $B$. Let us denote by $\delta_i$ the vanishing degree of the discriminant
 $\Delta$ over the divisor $D_i$. For higher degenerations this will also introduce 
non-trivial gauge-groups on the 7-branes. The precise group is encapsulated in the 
vanishing degree of the $a_i$ and $\Delta$~\cite{Bershadsky:1996nh}. 
For example, for  gauge group $G$ along the divisor $w=0$, where $w$ is one of the base 
coordinates $y_i$, the sections $a_n$ must take the following form:
\beq \label{TateSUn}
  a_1 = \mathfrak{b}_5 w^{\kappa_1} , \quad
  a_2 = \mathfrak{b}_4 w^{\kappa_2} , \quad
  a_3 = \mathfrak{b}_3 w^{\kappa_3} , \quad
  a_4 =  \mathfrak{b}_2 w^{\kappa_4} , \quad
  a_6 =  \mathfrak{b}_0 w^{\kappa_6} ,
\eeq
\beq
\begin{array}{c|c|ccccc}
G & E_8/G & \kappa_1 & \kappa_2 &\kappa_3 & \kappa_4 & \kappa_6\\ 
\hline
E_8 & SU(1) & 1 & 2 & 3 & 4 & 5\\
E_7 & SU(2) & 1 & 2 & 3 & 3 & 5\\
E_6 & SU(3) & 1 & 2 & 2 & 3 & 5\\
SO(10) & SU(4) & 1 & 1 & 2 & 3 & 5\\
SU(5) & SU(5) & 0 & 1 & 2 & 3 & 5 \\
SU(4) & SO(10) & 0 & 1 & 2 & 2 & 4
\end{array}\ 
\eeq
The sections $\mathfrak{b}_i$ generically depend on all coordinates $(y_i,w)$ 
of the base $B$ but do not contain an overall factor of $w$. It is important to stress 
that in case of such a higher degeneration not only the elliptic fibration will be 
singular, but rather the Calabi-Yau fourfold itself.

Let us specialise now to the Tate model of an SU(5) GUT theory along the divisor $S$ given by $w=0$. In our concrete geometry of section \ref{sec_GeomEx}, we therefore identify $w$ with the toric coordinate $y_3$.

Matter in the representations ${\bf 10}$ and ${\bf 5}$  is localised \cite{Friedman:1997yq} on curves on $S$ of further enhancement to $SO(10)$ and, respectively, $SU(6)$,
\bea \label{curve10}
  && P_{10}: \quad w=0 \quad \cap \quad \mathfrak{b}_5 = 0, \nonumber \\
  && P_5: \, \, \quad w=0 \quad \cap \quad P = \mathfrak{b}_3^2 \mathfrak{b}_4 - \mathfrak{b}_2 \mathfrak{b}_3 \mathfrak{b}_5 + \mathfrak{b}_0 \mathfrak{b}_5^2 = 0,
\eea
Note that the higher powers in $w$ appearing in the polynomials $\mathfrak{b}_i$ become irrelevant for the geometry of the matter curves. The same applies to the Yukawa couplings  ${\bf 10 \,  10 \, 5}$ and the ${\bf 10 \, \ov 5 \, \ov 5}$  due to their localisation on $S$. These are characterised by the simultaneous vanishing of the $w$-independent part of $\mathfrak{b}_i$ such as to produce point singularities of type $E_6$ and $SO(12)$, respectively.

\subsection{Spectral covers for $SU(5)$ models}

Motivated by this picture, a convenient way to describe the physics of a GUT model has been developed in \cite{Donagi:2008ca,Hayashi:2008ba,Hayashi:2009ge,Donagi:2009ra} in terms of the spectral cover construction \cite{Friedman:1997yq,donagi-1997-1} over $S$. While inspired by models with heterotic duals, it has proven a powerful method also in more general F-theory compactifications in particular to describe the gauge flux required for chirality. The general idea is to focus, for the description of the matter curves on $S$ and the gauge flux, on the neighbourhood of $S: w=0$ inside $B$ by restricting to the $w$-independent part of the sections $\mathfrak{b}_i$. The resulting polynomials will be called 
\bea
\label{bi}
b_i = \mathfrak{b}_i|_{w=0}
\eea
 in the sequel.

This structure is conveniently captured in a certain auxiliary, non-Calabi-Yau three-fold $X$ \cite{Donagi:2009ra}, which is a $\mathbb P^1$ bundle over the GUT divisor $S$.
The gauge group $G$ on $S$ is treated as the commutant of the group $H$ inside an underlying $E_8$. It is interpreted as the  result of breaking the maximal $E_8$ enhancement of the $\mathbb P_{1,2,3}[6]$ fiber by a non-trivial Higgs bundle \cite{Donagi:2008ca,Beasley:2008dc} with structure group $H$ over $S$. Part of this data is also to specify a gauge bundle over $S$ with structure group $H$.
The gauge flux is then treated in the same manner as bundles in heterotic compactifications on elliptic fibrations, except that one now works on a non-Calabi-Yau space and the generic fiber is a $\mathbb P^1$.  

Even though strictly speaking the spectral cover construction keeps track only of the information in the neighbourhood of $S$, its validity seems to go beyond the local limit.
As a non-trivial consistency check reference \cite{Blumenhagen:2009yv} has used the spectral cover construction to derive a general expression for the Euler characteristic for a certain class of F-theory models with non-abelian singularities. The fact that the resulting values agree in all tested cases where the Euler characteristic can be computed independently by toric geometry adds further credibility to the spectral cover approach.

In the remainder of this section we collect the main information on the spectral cover construction for the construction of $SU(5)$ GUT models. For more details see  \cite{Donagi:2008ca,Hayashi:2008ba,Hayashi:2009ge,Donagi:2009ra}.
The starting point is the auxiliary non-Calabi-Yau threefold $X$ \cite{Donagi:2009ra} constructed as a fibration over $S$, 
\beq \label{defX}
  X = {\mathbb P} ({\cal O}_{S} \oplus K_S) ,\qquad p_X: X \surjto S,
\eeq
where $p_X$ is the projection to the base of the bundle. The base $S$ is viewed as the vanishing locus of the section $\sigma$ in $X$ with self-intersection
\beq \label{sigma2}
  \sigma \cdot \sigma = - \sigma\,  c_1(S).
\eeq
 The first Chern class of $X$ is $c_1(X) = 2 \sigma + 2 c_1(S)$.

The spectral cover ${\mathcal C}^{(5)}$ of an SU(5) model is a 5-fold cover of $S$ inside this auxiliary threefold $X$. It is associated with Higgs bundle of structure group $H= E_8/SU(5) = SU(5)$. The physical significance of ${\cal C}^{(5)}$ is that its intersections with the ${\mathbb P}^1$ fiber represent the eigenvalues associated with the Higgs bundle \cite{Donagi:2008ca,Beasley:2008dc} of the GUT theory on $S$ \cite{Hayashi:2009ge,Donagi:2009ra}.
By slight simplification 
 ${\mathcal C}^{(5)}$ is given by 
\beq \label{C5b}
  b_0 s^5 + b_2 s^3 + b_3 s^2 + b_4 s + b_5 = 0,
\eeq
where $s=0$ corresponds to the base $S$ of $X$.\footnote{More precisely, $s=0$ denotes $S$ as the base of the total bundle $K_S$. The compactification of this total space is $X$. More details can be found in \cite{Donagi:2009ra}.}
The polynomials $b_i$ are indeed the $w$-independent part $(\ref{bi})$ of the sections $\mathfrak{b}_i$, as motivated above. They can therefore be viewed as sections entirely on $S$ since all information on the geometry normal to $S$ has been dropped by discarding the terms of higher order in $w$.
Their classes are
\beq \label{bisection}
  b_j \in H^0\big(S; {\cal O}(\eta - j c_1(S))\big) = H^0\big(S; {\cal O}((6 - j )c_1(S)  - t )\big)
\eeq
for a class
\beq
\eta = 6 c_1(S) - t, \quad\quad -t = c_1(N_{S/B})
\eeq
Since $\sigma$ is the class of $s=0$ in $X$ and with the assignment \eqref{bisection} for $b_i$, we have cohomologically in $X$
\beq
  [{{\cal C}^{(5)}}] =5 \sigma + \pi_5^* \eta,
\eeq
where  $\pi_5$ denotes the projection from the 5-fold cover  ${\mathcal C}^{(5)}$ onto $S$.

The intersection of ${\cal C}^{(5)}$ with $S$ yields the matter curve $P_{10}$ (\ref{curve10}) for the ${\bf 10}$ representation.
Generically  ${\cal C}^{(5)}$ is a connected divisor. In this case, the intersection locus with $S$ is also connected and all matter in the ${\bf 10}$ is localised on a single curve.
The $\bf 5$ curve $P_{5}$ is considerably more complicated and can be considered as the intersection of $S$ with another spectral cover ${\mathcal C}_{\wedge^2 V}$ associated with the anti-symmetric representation of $SU(5)$. What is important here is that the curve $P_5$ will also be connected for a generic connected spectral cover.

To avoid dimension 4 proton decay, however, it is necessary for the matter ${ \bf \ov 5}_m$ and the Higgs pair $[{\bf 5}_{H_u} + {\bf 5}_{H_d}]$ to be localised on two different curves \cite{Beasley:2008kw}. In fact the split has to be such that it
affects also the neighbourhood of $S$ inside $B$ \cite{Hayashi:2009ge}. Consequently the spectral cover itself has to factor in such a way that the curves $P_m$ and $P_H$ for ${ \bf \ov 5}_m$ and ${\bf 5}_H$ on $S$ split \cite{Hayashi:2009ge}.

\subsection{Split spectral covers}
\label{section-split}

In this paper we focus on the minimal split in agreement with absence of dimension 4 proton decay operators. That is we specify to 
the situation of a factorised divisor, analysed in detail in \cite{Marsano:2009gv}, of the form
\beq
  {\cal C}^{(5)} = {\cal C}^{(4)} \times {\cal C}^{(1)} ,
\eeq
corresponding to the factorisation of \eqref{C5b} into 
\beq \label{C4-1}
  (c_0 s^4 + c_1 s^3 + c_2 s^2 + c_3 s + c_4) (d_0 s + d_1) = 0.
\eeq
To avoid extra $\bf 10$ representations one takes 
 $d_1$ as an element of $H^0(S; {\cal O}_S)$, which is consistent with setting it to unity \cite{Marsano:2009gv}.
Comparison of \eqref{C4-1} with \eqref{C5b} allows one to express the sections $b_i$ as 
\beq \label{bcrel}
  b_5 = c_4, \quad b_4 = c_3 + c_4 d_0, \quad b_3 = c_2 + c_3 d_0, \quad b_2 = c_1 + c_2 d_0, \quad b_0 =  -c_1 d_0^2\ ,
\eeq
where we have further restricted ourselves to $ c_0 = -c_1 d_0$ such that the term proportional to $s^5$ vanishes in \eqref{C4-1}. 
This identifies the coefficients appearing in the factorised polynomials as sections
\beq \label{split_sections}
  \bal
    d_1 &\in H^0(X; {\cal O}), \qquad d_0 \in H^0(X; p_X^* (TS)), \\
    c_n &\in H^0(X; {\cal O}(p_X^*(\eta - (1+n) c_1(S))). 
  \eal
\eeq
The two components of ${\cal C}^{(5)}$ are in the respective classes
\beq
  [{{\cal C}^{(4)}}] = 4 \sigma + \pi_4^* \tilde \eta, \qquad [{{\cal C}^{(1)}}] =  \sigma + \pi_1^* c_1(S),
\eeq
where
\beq
\label{tildeeta}
  \tilde \eta = \eta - c_1(S), \qquad \eta = 6 c_1(S) + c_1(N_{S}).
\eeq
The matter curve $P_{10}$  on $S$ for representation ${\bf 10}$ is the projection of the curve  $[\ccP_{10}]  = {\cal C}^{(4)} \cap \sigma$ in $X$ to the section $\sigma$,
\beq
P_{\bf 10} =   [\ccP_{10}]|_{\sigma} =  \eta - 5 c_1(S) = \tilde \eta - 4 c_1(S).
\eeq
The computation of the matter curve for $\bf \ov 5_m$ and $\bf 5_H$ is much more complicated due to singularities in the associated antisymmetrised spectral cover \cite{Donagi:2004ia,Blumenhagen:2006wj, Hayashi:2008ba, Hayashi:2009ge}. Following the same logic as in \cite{Blumenhagen:2006wj} reference  \cite{Marsano:2009gv} finds for the present split 
\beq \label{classPH}
  \bal
    {}[\ccP_H] &=  2 \sigma \cdot \pi^*(2 \tilde\eta - 5 c_1(S)) + \pi^*(\tilde\eta - c_1(S)) \cdot \pi^*(\tilde\eta - 2 c_1(S)),  \\
    {}[\ccP_m] &=    \sigma \cdot \pi^*(  \tilde\eta - 2 c_1(S)) + \pi^*(c_1(S)) \cdot \pi^*(\tilde\eta - 2 c_1(S)).
  \eal
\eeq
The matter curves on $S$ follow by restricting intersection with $\sigma$ via (\ref{sigma2}).

\subsection{Gauge flux from spectral cover bundles}

In F-theory compactifications, gauge flux along 7-branes is described by 4-form flux $G_4$ with two legs along the brane.
Via duality with the heterotic string, the spectral cover approach translates this data into 
 a line bundle ${\cal N}$ on the spectral surface. Via the push-forward map onto $S$ this line bundle defines a vector bundle on $S$ with structure group $H= E_8/G$. As stated above, the gauge flux is thus formally treated as if it were responsible for the breaking of $E_8$ to the gauge group $G$ on $S$ as in the heterotic picture.
 
  For the above split of ${\cal C}^{(5)}$ into ${\cal C}^{(4)}$ and ${\cal C}^{(1)}$ \cite{Marsano:2009gv} this picture involves
two line bundles ${\cal N}_4$ and  ${\cal N}_1$ defined on the  respective components. The push-forward $\pi_{i*} {\cal N}_i$ defines a rank 4 bundle $V$ and, respectively, rank 1 bundle $L$ on $S$ such that the full bundle $W = V \oplus L$ has structure group $S[U(4) \times U(1)_X]$ \cite{Blumenhagen:2009yv}. The  $SU(5)$ gauge group realised on $S$ can be viewed as the non-abelian part of the commutant $SU(5) \times U(1)_X$ of the structure group of $V$ in $E_8$. The extra $U(1)_X$ factor becomes massive and is realised at best as a global selection rule.\footnote{This is a delicate issue. For non-zero VEVs of $U(1)_X$ charged fields  the gauge symmetry is higgsed and broken also as a selection rule. In the present case the corresponding recombination moduli of the two spectral surfaces  are states localised away from $S$ and are, strictly speaking, beyond the reliable scope of the spectral cover approach.}

For our purposes it suffices to restrict ourselves to so-called universal gauge flux  \cite{Donagi:2009ra}, which can be switched on for generic complex structure moduli of the (split) spectral cover.
In analogy to the definition of $U(n)$ bundles in the heterotic context \cite{Andreas:2004ja,Blumenhagen:2006wj}, a convenient parametrisation for $c_1({\cal N}_4)$ is \cite{Blumenhagen:2009yv}
\beq \label{linebundlesu4}
  \bal
    c_1(\mathcal{N}^{(4)})&= \frac{r^{(4)}}{2} + \gamma^{(4)}_u + \frac{1}{4} \pi^*_4 \zeta \\
    &= \left({\textstyle  1+4\lambda }\right)\,\sigma +   
       \left({\textstyle \frac{1}{2} - \lambda}\right) \pi_4^\ast \tilde\eta + \left( {\textstyle - \frac{1}{2}+ 4
       \lambda} \right) \pi_4^\ast c_1(S) + {\textstyle \frac{1}{4}}\, \pi_4^\ast \zeta,
  \eal
  \eeq
while for ${\cal N}_1$ one can formally set 
\beq
  c_1({\cal N}_1) =  - \pi^*_1 \zeta. 
\eeq
Here $\zeta$ is a class in $H^2(S, \mathbb Z)$.
Indeed, the above choice guarantees that $c_1(V) + c_1(L) =0$ \cite{Blumenhagen:2009yv}.
The flux is subject to a quantisation condition which ensures
integrality of $c_1({\cal N}_4)$ and $c_1({\cal N}_1)$, 
\beq \label{Nintegerb}
    4 \,\lambda  \in {\mathbb Z}\ , \qquad 
    \left({\textstyle \frac{1}{2}-\lambda}\right)\,\tilde \eta - {\textstyle \frac{1}{2}} c_1(S) + {\textstyle \frac{1}{4}}\, \zeta   \in  H^2(S; {\mathbb Z})\ .
\eeq

The purpose of switching on this gauge flux is to achieve a chiral matter spectrum localised on the described curves on $S$. The resulting chirality of the ${\bf 10}$,  ${\bf \ov 5_m}$ and ${\bf  5_H}$ fields can be computed in close analogy to the corresponding expressions for heterotic spectral covers.
Here we merely list the final results from \cite{Blumenhagen:2009yv} in terms of the above parametrisation of the spectral cover line bundle, 
\bea \label{chi10-split}
  &&  \chi_{\bf 10} = \chi(\ccP_{10}, {\cal N}_4 \otimes K_S |_{\ccP_{10}} ) 
     = \big( - \lambda \tilde \eta + {\textstyle \frac14} \zeta \big) \, (\tilde \eta - 4 c_1(S)),  \nonumber \\
 &&     \chi_{{\bf \ov 5}_m} = \chi ({\ccP}_m, {\cal N}_4  \otimes  {\cal N}_1  \otimes K_S |_  {{\ccP}_m}  ) 
    = \lambda \left( -\tilde\eta^2 + 6\tilde\eta c_1(S) -8 c^2_1(S)\right) + \textstyle \frac{1}{4} \zeta\, ( -3\tilde\eta +6 c_1(S)), \nonumber  \\
   &&   \chi_{{\bf \ov 5}_H} = 
  \lambda \left( -2\tilde\eta c_1(S) +8 c^2_1(S)\right) + \textstyle \frac{1}{4} \zeta\, ( 4\tilde\eta -10  c_1(S)).
\eea
Here all intersections are taken directly on $S$.

Finally, the breaking of $SU(5)$ to the Standard Model gauge group is achieved by switching on a line bundle ${\cal L}_Y$ on $S$ corresponding to  hypercharge flux \cite{Beasley:2008kw, Donagi:2008kj}. A consistent definition of this flux in a way that does not lead exotic states from the decomposition of the $\bf 24$ representation of $SU(5)$ has been given in \cite{Blumenhagen:2009yv}. There the GUT bundles are twisted by fractional powers of ${\cal L}_Y$ in a manner respecting the subtle quantisation conditions for the gauge flux.

\subsection{Supersymmetry and D3-tadpole}
\label{sec_D3}

The above flux is subject to two further consistency conditions.
The first is the well familiar D-term supersymmetry condition, which requires the Fayet-Iliopoulos term of $U(1)_X$
\beq
\label{FIterm}
  \mu({V}) = \int_S  {\iota}^*J \wedge \zeta  = - \mu({L}) ,
\eeq
to vanish in absence of VEVs for matter charged under $U(1)_X$. This condition has to hold inside the K\"ahler cone. Note that if one aims at realising the decoupling limit described in section \ref{Decoupling}, care has to be taken whether the resulting constraints are compatible with (\ref{FIterm}).

The second, important constraint stems from the necessity of 
cancelling the induced D3-brane tadpole in compact string vacua as already given in 
\eqref{chi/24}.
To avoid uncontrolled supersymmetry breaking by anti D3-branes, the flux contribution must not overshoot the curvature part
induced by the Euler characteristic $\chi(Y)$.
For a single $U(n)$ spectral cover bundle the flux piece takes the form \cite{Curio:1998bva, Andreas:2004ja}
 \bea
 \label{piga}
N_{flux} =  \frac12 \int_{Y} G \wedge G =   \frac{1}{2}  \int_S \left(  \lambda^2 n  \eta(\eta - n c_1(S)) - \frac{1}{n} \zeta^2      \right).
\eea
For the present 4-1 split spectral cover, we have an $S[U(4) \times U(1)]$ bundle, which contributes  \cite{Blumenhagen:2009yv} 
\beq
\label{flux4-1}
N_{flux} =  \frac{1}{2}  \int_S \left(  4 \lambda^2   \tilde\eta( \tilde\eta - 4 c_1(S)) - \left(\frac{1}{4} +1\right) \zeta^2  \right),
\eeq
with $\tilde \eta$ as in (\ref{tildeeta}).
Switching on hypercharge flux ${\cal L}_Y$ on $S$ to accomplish GUT symmetry breaking  modifies the full flux induced D3-charge further by an extra $ - c_1^2({\cal L}_Y) + \zeta c_1({\cal L}_Y) $ on the right-hand side of (\ref{flux4-1}).

Crucially, the value of  $\chi(Y)$ entering the D3-brane tadpole keeps track of the presence of non-abelian gauge groups on $Y$.  The degenerations of the elliptic fiber over the locus of non-abelian 7-branes render the Calabi-Yau $Y$ singular. The value appearing in (\ref{chi/24}) refers to the smooth fourfold obtained by resolving these singularities. This reduces the value for $\chi(Y)$  as compared to the naive value obtained by considering merely a smooth Weierstrass model over the base $B$ with at worst $I_1$ singular fibers. 

In \cite{Blumenhagen:2009yv}  a general expression for the Euler characteristic of an F-theory model that can be described by a spectral cover with structure group $G$ has been given,\footnote{An alternative method to directly compute $\chi$ of the resolved space is pursued in \cite{Andreas:1999ng,Andreas:2009uf}.} 
\beq \label{chi-prop}
  \chi(Y) = \chi^*(Y) + \chi_{G} - \chi_{E_8}.
\eeq
Here $\chi^*(Y)$ is the Euler characteristic for the elliptic fibration over $B$
in the absence of the resolved gauge group. In our cases removing the gauge singularity leads to 
a fourfold for which $\chi^*(Y)$ is simply computed via \eqref{smooth_chi}.  
The quantities $\chi_{G}$, $\chi_{E_8}$ are defined in terms of the class $\eta$ defining the spectral cover. Since we are interested in this paper in an $S[U(4) \times U(1)_X]$ spectral cover, we only display the expression for $G=SU(n)$ and for $E_8$, referring the reader to \cite{Blumenhagen:2009yv} for further details:
\bea
\label{chi_SCC}
\chi_{E_8} & =& 120 \int_S  \bigl( 3\eta^2 - 27\eta c_1(S) + 62 c_1^2(S)\bigr), \\
\chi_{SU(n)} &=&  \int_S c_1^2(S) (n^3-n) +3n\, \eta \big(\eta-n c_1(S)\big).
\eea
This simple proposal for the computation of the Euler characteristic has been checked in \cite{Blumenhagen:2009yv}  in a number of torically realised examples, which thus admit the independent computation of $\chi(Y)$ after resolution of the non-abelian singularities. 
In the context of a split spectral cover, the factor $\chi_G$ is to be replaced by the sum $\chi_{SU(4)} + \chi_{SU(1)}$, where the latter vanishes trivially and the $SU(4)$ part involves the class $\tilde \eta = \eta - c_1(S)$.

\section{Explicit 3-Generation Example} \label{sec_3Gen}

We are now in a position to provide the details for the construction of D3-tadpole cancelling, compact $SU(5)$ GUT models based on a split ${\cal C}^{(4)} - {\cal C}^{(1)}$ cover. Let us present one explicit example realised within the particular geometry described in section \ref{sec_GeomEx}. In appendix \ref{sec_moresol1} we  collect the data of a sample of extra solutions we found. A comparable construction is possible also on the geometry described in appendix \ref{appExample2}, leading likewise to a number of flux three-generation solutions in agreement with the D3-tadpole.

\subsubsection*{Matter curves on $S$}

As a first step we determine the matter curves on the GUT divisor $S$ given by $y_3=0$.
The general formulae for these curves (as objects in the auxiliary spectral cover threefold $X$) are given in section \ref{section-split}.  With the concrete expressions for $c_1(B)$, $c_1(S)$, $c_1(N_S)$ in the geometry of section \ref{sec_GeomEx} one finds for the matter curve classes on $S$
\beq
  \bal
  {}  [P_{\bf 10}] &= c_1(B) = (S + Y_1 + Y_2)|_S ,\\
    [P_{\bf 5_H}] &= 5 c_1(S) + 2 c_1(N_S) = (2S + 5 Y_1 + 5 Y_2)|_S ,\\
    [P_{\bf \ov 5_m}] &= 3 c_1(S) + c_1(N_S) = (S + 3 Y_1 + 3 Y_2)|_S
  \eal
\eeq\\
in terms of the pullback of divisor classes of $B$.
With the help of the explicit pullback map (\ref{pullback1}),  (\ref{pullback2})  for the individual classes this leads to the following expression for the matter curve classes in $H_2(S, \mathbb Z)$, 
\beq
[P_{\bf 10}] = 2 l - E_I, \qquad [P_{\bf 5_H}] = 13 l - 3 E_1 - 5 E_I, \qquad [P_{\bf \ov 5_m}] = 8 l - 2 E_1 - 3 E_I .
\eeq
Now, we can switch on hypercharge flux along, say, the element
\beq
c_1(\cL_Y) = E_3 - E_4.
\eeq
Since this class is cohomologically trivial on the ambient base $B$, this flux breaks $SU(5)$ to $SU(3) \times SU(2) \times U(1)_Y$ while keeping  $U(1)_Y$  massless. 
Our choice of ${\cal L}_Y$ guarantees that it restricts trivially to $[P_{\bf 10}]$, $[P_{\bf \ov 5_m}]$ and  $[P_{\bf 5_H}] $. Thus GUT symmetry breaking is realised in such a way that the GUT representations descend to an equal number of MSSM representations. While this is welcome for ${\bf 10}$ and ${\bf \ov 5_m}$, the Higgs doublet-triplet splitting must be solved here via non-trivial Wilson lines as described in the IIB context in detail in \cite{Blumenhagen:2008zz}.

While the main scope of this short article is to demonstrate that the global model building constraints can be met within a well-defined geometry, we note that the localisation of the Higgs field on a single curve requires fine tuning to avoid a large $\mu$-term \cite{Beasley:2008kw}.  
This can be avoided if the Higgs curve splits further into two single components for $H_u$ and $H_d$ in such a way that ${\cal L}_Y$ restricts to $+1$ and $-1$, respectively \cite{Beasley:2008kw}. On such a split curve the triplet  is automatically projected out and a single $H_u-H_d$ pair is kept.

It is in principle possible to further split $[P_{5_H}]$ to $[P_{5_{H_u}}]$ and $[P_{5_{H_d}}]$ consistent with this requirement, e.g. 
\beq
  \bal
    {} [P_{5_{H_u}}] = 7 l - 2 E_1 - 3 (E_2 + E_3) - 2 (E_4 + E_5),\\
    [P_{5_{H_d}}] = 6 l - 1 E_1 - 2 (E_2 + E_3) - 3 (E_4 + E_5).
  \eal
\eeq
However, thus far no explicit implementation of this in the coordinates $\{y_i\}$ has been found.
We plan to return to this technicality in the future. We also note that in the present setup a closer analysis of the Higgs triplet is required in view of gauge coupling unification \cite{Blumenhagen:2008aw}. The thresholds of the triplets have the potential to correct the hyperflux induced deviation from exact coupling unification at the GUT scale without the need to introduce extra thresholds. Models with extra incomplete GUT multiplets to achieve gauge unification are considered in \cite{Marsano:2009wr} (see also \cite{Leontaris:2009wi}).

\subsubsection*{Fixing the gauge flux to achieve 3 chiral generations}

We now construct consistent spectral cover line bundles ${\cal N}_4$ and  ${\cal N}_1$ which lead to three chiral generations of Standard Model matter (and no exotic representations).
For this purpose we go back to the parametrisation (\ref{linebundlesu4}) and make the ansatz $\lambda = \frac{x}{4}, x\in\mathbb{Z}$ in agreement with the quantisation condition (\ref{Nintegerb}). The second  constraint therein now becomes $\zeta = -x B + 2 X + 4 \tilde{\zeta}$ for $\tilde{\zeta}= m_0 X + m_i Y_i$ with $m_0, m_i \in \mathbb Z$. A search for $m_i$ such that $\chi_{5_H} = 0$ and $\chi_{10} = \pm3$ yields several  viable solutions, amongst which $\vec{m} = (1,0,3,0)$ is one of the more convenient ones. It leads to $x=-3$ and therefore to the following expression for $\zeta$,
\begin{equation}
\zeta = (3X - 3Y_1 + 9Y_2)|_S, \qquad \lambda = -\frac{3}{4} .
\end{equation}\\
In this case $\chi_{10} = +3$, establishing indeed the presence of three chiral generations of GUT matter.

\subsubsection*{D-Term supersymmetry and decoupling}
With the K\"ahler form $J= \sum_i r_i K_i$ expanded in terms of the generators (\ref{Kcone}) of the K\"ahler cone and $\zeta$ defined as above, the Fayet-Iliopoulos term (\ref{FIterm}) becomes
\beq
  \mu(V) = \int_S  \iota^*J \wedge \zeta = 6 r_1 - 12 r_2 + 12 r_4.
\eeq
Indeed  $\mu(V)$ can be arranged to vanish inside the K\"ahler cone. Most importantly, a vanishing Fayet-Iliopoulos is not in conflict with the physical decoupling limit described in section \ref{Decoupling} as the above term does not depend on $r_3$. Recall from section \ref{Decoupling} that it is the limit  $r_3 \rightarrow \infty$ and all other $r_i$ finite which controls $M_{Pl.} \rightarrow \infty$ while $M_{GUT}$ finite.  In more physical terms, we have found  
 a supersymmetric solution which is consistent with engineering the mild hierarchy $M_{GUT} \simeq 10^{-3} M_{Pl}$.

\subsubsection*{3-brane Tadpole}

To compute the curvature induced D3-brane charge for $B$ we follow the logic spelled out in section \ref{sec_D3} and first compute the bare Euler characteristic (\ref{smooth_chi}) for a \emph{smooth} fibration over our base $B$. From $c_1(B) = S + Y_1 + Y_2$ and $c_2(B) $ as given in appendix \ref{appExample} this is $\chi^*(Y)/24 = 282$. 
Next, the expression (\ref{chi_SCC}) for $\chi(E_8)$ must be computed from $\eta= 6 (Y_1 + Y_2) + S$, while for the split spectral cover $\chi(SU(4))$ is defined merely in terms of $\tilde \eta = 5 (Y_1 + Y_2) + S$.
In total the full curvature induced D3-brane tadpole of our geometry is
\beq
\frac{\chi(Y)}{24} = 226.
\eeq
Furthermore, the flux-dependent part  (\ref{flux4-1}) for our above choice of $\zeta$ gives
\beq
N_{flux} = 144.
\eeq

Putting everything together and remembering  that the hyperflux contributes with  $c_1^2(\cL_Y) = -2$ (while $c_1(\cL_Y) \, \zeta = 0$), we thus find that D3-tadpole cancellation requires the addition of
\beq
n_3 = 226 - 144 - 2 = 80
\eeq
D3-branes.

\section{Conclusions}\seclabel{sec_concl}
In this note we have 
constructed several compact three-generation $SU(5)$ GUT models in the framework of F-theory, 
building on and extending the analysis in \cite{Blumenhagen:2009yv}. 
We have focused on two outstanding technical challenges. First our models allow for D3-brane tadpole cancellation without anti-D3-branes thanks to a sufficiently high Euler characteristic of the Calabi-Yau fourfold. Second we have demonstrated the possibility of decoupling gravity in situations  where the GUT brane is contractible only to curves, as is the case here.

Our approach of constructing the elliptic compactification fourfold  via toric methods is more general and by no means restricted to the specific examples presented in this paper. 
As we have stressed several times the machinery of toric geometry keeps control of the full compact Calabi-Yau fourfold and the singularities encoding the non-abelian gauge symmetry. This fact  enables us e.g.\ to compute the Euler characteristic of the fourfold, which is an important step in the investigation of D3-tadpole cancellation.
Concretely, we start with an elliptic fibration over a Fano threefold, defined as a complete intersection in a toric ambient space. In a next step a singularity generated on this base is blown up to a non-generic del Pezzo surface. One of the divisors obtained in this manner defines our GUT seven-brane. 
Note that these transitions generically render the base  non-Fano and toric methods become an important tool 
to guarantee the existence of the fourfold. The construction of this model  then allows us to 
find an explicit implementation of the decoupling limit $M_{Pl} \rightarrow \infty$, demonstrating 
the physical viability of situations where  the GUT divisor is contractible to a curve instead of to a point.

Using the spectral cover approach to describe the gauge flux necessary for chiral matter, we have furthermore obtained a number of correctly quantised flux solutions whose 3-brane charge does not exceed that of the manifold. Moreover the Fayet-Iliopoulos term in the supersymmetry condition can be arranged to vanish without affecting the physical decoupling limit.

 We have provided two example manifolds as well as, in each case, several solutions for the gauge flux exhibiting the properties described above. The chiral spectrum in these models is that of the Standard Model.
Even though a detailed phenomenological analysis is beyond the main focus of this article, we point out that a property that has thus far proven difficult to realise in detail is the consistent split of the $5_H$-matter curve. This is required in order to avoid dimension-five proton decay and the $\mu$-problem. While in principle such a split can be given in terms of the matter curve classes expressed as elements of the homology of the GUT-brane, an explicit implementation in the ambient space coordinates is yet to be achieved. We hope to return to this question in the future.



\subsection*{Acknowledgements}
We gratefully acknowledge discussions with R. Blumenhagen, A. Braun, C. Cordova, T.-W. Ha, A. Hebecker, B. Jurke, A. Klemm and D. Klevers.
TG and TW would like to thank  the MPI Munich for hospitality during the preparation of this work. 
This work was supported in parts by the 
SFB-Transregio 33 ``The Dark Universe'' by the DFG and the Klaus-Tschira-Foundation.

\appendix
\newpage
\section{On the toric construction of elliptic fourfolds}\label{appToric}

In this appendix we describe the details of the construction of an elliptic Calabi-Yau space $Y$ as a complete intersection within a toric ambient space $\mathbb P_{\Delta}$ specified in \eqref{PDelta}.
A powerful way to formulate the geometry of $Y$ and its ambient space is via
reflexive polyhedra. Essentially, these consist of integral vectors which are related 
via the $\ell^k$. We refer the reader to ref.~\cite{Cox:2000vi} for an introduction on this 
subject. Let us denote by $\Delta^*$ the polyhedron in which the points correspond 
to the coordinates $x_i$, or divisors $D_i$, of $\bbP_{\Delta}$. The split 
in \eqref{nef-partition} is encoded by a split of $\nabla = (\nabla_1,\nabla_2, \ldots )$ in 
partitions $\nabla_a$. These data can encode a non-singular complete intersection 
if $\nabla$ and the Minkowski sum $\nabla_1 + \nabla_2 + \ldots$ are both reflexive polyhedra.\footnote{We have been
a bit sloppy in this treatment, since one has to add the origin to each $\nabla_i$. 
The set $\nabla_1 + \nabla_2$ is then defined to consist of the sums of each point in $\nabla_1$ with each point in $\nabla_2$. A polyhedron is reflexive if the origin is the only interior point of the polyhedron.} In case this condition is satisfied one calls the split $(\nabla_1,\nabla_2,\ldots )$ a nef-partition \cite{Borisov-1993, Kreuzer:2001fu}. 
To explicitly describe the polynomials $p_a(x_1,\ldots,x_n)$ defining $Y$ we first 
have to introduce the dual Newton polyhedra $\Delta = (\Delta_1,\Delta_2,\ldots)$ via~\cite{Cox:2000vi}
\beq \label{NablaDelta_dual}
  \langle \nabla_a ,\Delta_b \rangle \geq - \delta_{ab} .
\eeq 
While the $(\nabla_1,\nabla_2,\ldots)$ correspond to coordinates and toric divisors for the complete intersection, 
the points in $(\Delta_1,\Delta_2,\ldots)$ correspond to the monomials of the constraints $p_a=0$. More precisely, the complete intersection is given by the constraints 
\beq \label{complete_int}
  p_a = \sum_{w_k \in \Delta_a} c^{(a)}_{k} \prod_{b} \prod_{\nu_i \in \nabla_b} x_i^{\langle \nu_i,w_k \rangle+\delta_{ab}}= 0 \ .
\eeq
The coefficients $\smash{c_k^{(a)}}$ are a redundant parametrisation of the complex structure deformations of $Y$.
While this general formalism may seem rather abstract it turns out to be very tractable when applied 
to concrete examples.

The toric constructions are readily carried out for elliptically 
fibered Calabi-Yau fourfolds with base spaces \eqref{small_list} and 
$\bbP_{123}[6]$ elliptic fiber. Let us focus here on the example with 
$B = \bbP^4[3]$. The toric data for the nef-partitions $(\nabla_1,\nabla_2)$ 
are listed in Table \ref{exampleData_P3}.
\begin{table}[ht]
  \centering
  \begin{tabular}{c|r@{\,$=$\,(\,}r@{,\;\;}r@{,\;\;}r@{,\;\;}r@{,\;\;}r@{,\;\;}r@{\,)\;\;}|c} 
    nef-part. &\multicolumn{7}{c|}{vertices} & coords.  \\
    \hline\hline
    $\nabla_1$ & $\nu_1$ & $-1$ & $0$ & $0$ & $0$ & $0$ & $0$ & $y$$^\big.$   \\
    & $\nu_2$ & $0$ & $-1$ & $0$ & $0$ & $0$ & $0$ & $x$   \\
    & $\nu_3$ & $3$ & $2$ & $0$ & $0$ & $0$ & $0$ & $z$  \\
    & $\nu_4$ & $3$ & $2$ & $-1$ & $-1$ & $-1$ & $-1$ & $y_1$  \\
    & $\nu_5$ & $3$ & $2$ & $1$ & $0$ & $0$ & $0$ & $y_2$ \\
    \hline
    $\nabla_2$ & $\tilde \nu_1$ &   $0$  &   $0$  &   $0$  &   $1$  & $0$ & $0$ & $y_3$$^\big.$ \\
    & $\nu_7$ &   $0$  &   $0$  &   $0$  &   $0$  &   $1$  &   $0$ & $y_4$ \\
    & $\nu_8$ &   $0$  &   $0$  &   $0$  &   $0$  &   $0$  &   $1$ & $y_5$ \\
    \hline
  \end{tabular}
  \caption{\small Toric data for the elliptic Calabi-Yau fourfold over the base $\bbP^4[3]$.}
  \label{exampleData_P3}
\end{table}
One notes that the polyhedron for $\bbP_{123}$, spanned by the three toric points $\tilde \nabla=((-1,0),(0,-1),(3,2))$,
appears in the first two columns of each $\nabla_1$. The complete intersection is 
elliptically fibered since $\nabla$ contains the 
points $\nu_1,\nu_2,\nu_3$, corresponding to the 
coordinates $(y,x,z)$ of the elliptic fiber, which have the points $\tilde \nabla$ 
in the first two entries but are zero otherwise~\cite{Avram:1996pj}. 
The polyhedron of the base $B$ appears in the last four rows of $\nabla$. 
It is not hard to check that the splits of $\nabla$ into $(\nabla_1,\nabla_2)$
determine a valid nef-partition.\footnote{This can be done by using the program PALP \cite{Kreuzer:2002uu}.}
Since $\nabla_1$ contains the points $(\nu_1,\nu_2,\nu_3)$ corresponding 
to the coordinates $(x,y,z)$ of the elliptic fiber one 
checks using \eqref{complete_int} that $p_1 \equiv P_{\rm W}=0$ is precisely the Tate form \eqref{Tate1}. 
The coefficients $a_n$ in the Tate form can be explicitly given in terms of the toric data. Let us first introduce the sets 
\beq \label{def-Ar}
  A_r = \{w_k \in \Delta_1:\, \langle \nu_3,w_k \rangle = r+1\} .
\eeq 
These are the elements in the Newton polyhedron which generate the monomials in the Tate form $p_1=0$ containing the power $z^r$. Hence, we can write 
\beq \label{toric_ai}
  a_r = \sum_{w_k \in A_r} c^{(1)}_{k} \prod_{n=1}^2 \prod_{\ \nu_i \in \nabla_n,\,i>3} \!\!\!\! y_i^{\langle \nu_i,w_k \rangle+\delta_{mn}} ,
\eeq
where we recall that $\nu_3$ corresponds to the $z$-coordinate of the elliptic fiber in \eqref{Tate1} and hence $a_r$ appears in front of $z^r$. Moreover, $p_2=0$ is the constraint for the base $B$, i.e.~for the 
example in Table \ref{exampleData_P3} one finds $\bbP^{4}[3]$. 

The del Pezzo surfaces supporting the GUT brane can now be obtained by adding further toric 
points to the nef-partitions in Table \ref{exampleData_P3}. In particular, one notes that 
such a construction blows up a singular curve by adding the point $(3,2,-1,-1,0,0)$ 
to the first nef-partition $\nabla_1$. One checks that this blows up 
a del Pezzo $4$ surface. The fact that this blow-up arises from a singular curve 
can be understood torically as follows. If one projects the polyhedron 
of Table \ref{exampleData_P3} to the base $B$ by omitting the first two columns, 
then the new del Pezzo divisor $(-1,-1,0,0)$ subdivides the two-dimensional cone
spanned by $(1,0,0,0)$ and $(0,1,0,0)$ in the polyhedron of $B$. Upon
adding this point the curve corresponding to this cone is
removed from $B$ and replaced by the new del Pezzo divisor. In this
way we will systematically proceed to construct the desired 
Calabi-Yau fourfold (see Appendix \ref{appExample} for the toric details).

\subsubsection*{Resolving singularities in elliptic Calabi-Yau fourfolds}\seclabel{toric_resolution_ex}
Having constructed the Calabi-Yau fourfold as an elliptic fibration over the base $B$ with a del Pezzo surface, we next want to use toric geometry to degenerate the elliptic fiber of this surface to an $SU(5)$ singularity~\cite{Candelas:1996su, Candelas:1997eh}. Our strategy is to use the explicit expressions for the $a_i$ given in \eqref{toric_ai} and drop all monomials that would violate the $SU(5)$ form \eqref{TateSUn}. Recall that in  \eqref{TateSUn} we have demanded that the $(a_1,a_2,a_3,a_4,a_6)$ contain the overall factors $(1,w,w^2,w^3,w^5)$ with coefficient functions $\mathfrak{b}_r(w,y_i)$. Hence, one has to drop the monomials in $a_r$ which admit powers of $w^k$ with $k<(0,1,2,3,5)$, respectively. 
Torically this is achieved by dropping points in the Newton polyhedron $\Delta_1$. One thus drops points of the sets $A_r$, defined in \eqref{def-Ar}, encoding the monomials in $a_r$, and obtains new sets  
\beq
  A_r^{SU(5)} \subset A_r  , \qquad \Delta_1^{SU(5)} \subset \Delta_1 . 
\eeq
Using the new Newton polyhedron in the Tate form $p_1 = 0$ in \eqref{complete_int} ensures that $Y$ will admit the singularity of the desired $SU(5)$ type. Clearly, one can use this method to generate also higher gauge groups on the divisor $S$ by imposing stronger constraints on the allowed monomials. Maximally, we can degenerate the elliptic fiber to $E_8$ by reducing to $\Delta_1^{E_8}$ dropping all monomials in the $a_r$ with powers lower than $(w,w^2,w^3,w^4,w^5)$.

Toric geometry can now be used to automatically resolve the singularities of the elliptic fibration. In order to do that one has to add new blow-up divisors, or, equivalently, new points to the polyhedron $\nabla=(\nabla_1,\nabla_2)$. More precisely, this can be done by determining the new duals of $(\Delta^{SU(5)}_1,\Delta_2^{SU(5)} = \Delta_2)$ via
\beq \label{NablaDelta_dualSU(5)}
  \langle \nabla^{SU(5)}_n ,\Delta^{SU(5)}_m \rangle \geq - \delta_{mn} 
\eeq 
just as in \eqref{NablaDelta_dual}. Since $\Delta^{SU(5)}\subset \Delta$ the polyhedron $\nabla^{SU(5)}=(\nabla^{SU(5)}_1,\nabla^{SU(5)}_2)$ will contain more points than the original polyhedron $\nabla$. The extra points $\tilde \nu_i$ are in $\smash{\nabla_1^{SU(5)}}$ and explicitly given by
\beq \label{blowup}
  \begin{array}{rr@{\,=\,(\,}r@{,\;\;}r@{,\;\;}r@{\,)\;\;}}
    \nabla_1^{SU(5)}: & \nu_{dP}       & 3 & 2 & \vec{\mu}, \\
                      & \tilde\nu_1 \ & 2 & 1 & \vec{\mu}, \\
                      & \tilde\nu_2 \ & 1 & 1 & \vec{\mu}, \\
                      & \tilde\nu_3 \ & 1 & 0 & \vec{\mu}, \\
                      & \tilde\nu_4 \ & \;0 & \;0 & \vec{\mu},
  \end{array}
\eeq
where the point $\nu_{dP}$ corresponds to the GUT del Pezzo divisor in the base. 
The new points \eqref{blowup} together with the original points 
define a new complete-intersecting fourfold $Y^{SU(5)}$, which admits a resolved $SU(5)$ fiber. 
Using these data one can again determine all relevant topological data for the fourfold 
such as intersection numbers and Chern classes. Clearly, this allows us to 
directly compute $\chi(Y^{SU(5)})$. 

Let us stress that one can use this technique to generate and resolve singularities also for other gauge groups realised on $S$ or some other toric divisor. In particular, one finds for the maximal $E_8$ case that one has to add the points $(3,2,n \vec{\mu}),\, n=1,...,6$, $(2,1,n \vec{\mu}),\, n=1,...,4$, $(1,1,n \vec{\mu}), n=1,2,3$, $(1,0,n \vec{\mu}), n=1,2$ and $(0,0,\vec{\mu})$.

\section{Details of a 3-generation model based on $\mathbb P^4[3]$} \label{appExample}

\subsection{The fourfold}

In this appendix we provide the toric data and extra geometric details of the fourfold introduced in section \ref{sec_GeomEx} on which we construct consistent three-generation $SU(5)$ models. We begin with the toric data for the model as given in \tableref{exampleData}.
\begin{table}[ht]
  \centering
  \begin{tabular}{c|r@{\,$=$\,(\,}r@{,\;\;}r@{,\;\;}r@{,\;\;}r@{,\;\;}r@{,\;\;}r@{\,)\;\;}|c|c@{\ \;\;}c@{\ \;\;}c@{\ \;\;}c@{\ \;\;}c@{\ \;\;}} 
    nef-part. &\multicolumn{7}{c|}{vertices} & coords. & \multicolumn{5}{c}{U(1)-charges}$_\big.$ \\
    \hline\hline
    & $N_0$ & $0$ & $0$ & $0$ & $0$ & $0$ & $0$ & $x_0$$^\big._\big.$ & $B$ & $S$ & $Y_1$ & $Y_2$ & $Y_3$\\
    \hline
    $\nabla_1$ & $\nu_1$ & $-1$ & $0$ & $0$ & $0$ & $0$ & $0$ & $y$$^\big.$ & $3$ & $3$ & $3$ & $3$ & $\cdot$ \\
    & $\nu_2$ & $0$ & $-1$ & $0$ & $0$ & $0$ & $0$ & $x$  & $2$ & $2$ & $2$ & $2$ & $\cdot$ \\
    & $\nu_3$ & $3$ & $2$ & $0$ & $0$ & $0$ & $0$ & $z$  & $1$ & $\cdot$ & $\cdot$ & $\cdot$ & $\cdot$ \\
    & $\nu_4$ & $3$ & $2$ & $-1$ & $-1$ & $-1$ & $-1$ & $y_1$  & $\cdot$ & $\cdot$ & $1$ & $\cdot$ & $\cdot$\\
    & $\nu_5$ & $3$ & $2$ & $1$ & $0$ & $0$ & $0$ & $y_2$ & $\cdot$ & $\cdot$ & $\cdot$ & $1$ & $\cdot$\\
    & $\nu_6$ & $3$ & $2$ & $-1$ & $-1$ & $0$ & $0$ & $y_3$$_\big.$  & $\cdot$ & $1$ & $\cdot$ & $\cdot$ & $\cdot$\\
    \hline
    $\nabla_2$ & $\tilde \nu_1$ &   $0$  &   $0$  &   $0$  &   $1$  & $0$ & $0$ & $y_4$$^\big.$ & $\cdot$ & $1$ & $1$ & $\cdot$ & $\cdot$\\
    & $\nu_7$ &   $0$  &   $0$  &   $0$  &   $0$  &   $1$  &   $0$ & $y_5$ & $\cdot$ & $\cdot$ & $1$ & $\cdot$ & $1$\\
    & $\nu_8$ &   $0$  &   $0$  &   $0$  &   $0$  &   $0$  &   $1$ & $y_6$ & $\cdot$ & $\cdot$ & $1$ & $\cdot$ & $\cdot$\\
    & $\nu_9$ &   $0$  &   $0$  &   $-1$  &   $0$  &   $0$ &    $0$ & $y_7$ & $\cdot$ & $-1$ & $-1$ & $1$ & $-1$\\
    & $\nu_{10}$ &   $0$  &   $0$  &   $-1$  &   $0$  &   $-1$ &    $0$ & $y_8$$_\big.$ & $\cdot$ & $\cdot$ & $\cdot$ & $\cdot$ & $1$\\
    \hline
  \end{tabular}
  \caption{\small Toric data for the elliptic Calabi-Yau fourfold of section \ref{sec_GeomEx}.}
  \tablelabel{exampleData}
\end{table}\\
Upon expanding the total Chern class of the basis for this model one obtains
\beq
  \bal
    c_1(B) = &\ S + Y_1 + Y_2,\\
    c_2(B) = &\ -S^2 - S Y_1 + 2 S Y_2 - S Y_3 + 4 Y_1 Y_2 + 2 Y_2 Y_3 - Y_3^2,\\
    c_3(B) = &\ S^2 (-S - 3Y_1 + Y_2 - 2 Y_3) + S (-2 Y_1^2 + 5 Y_1 Y_2 - 3 Y_1 Y_3 + 3 Y_2 Y_3 - 2 Y_3^2) \\
                 &+ Y_1 (-2Y_1^2 - 4 Y_1 Y_3 - 3 Y_3^2) - Y_2 Y_3^2.
  \eal
\eeq
The non-zero intersection numbers are given in \eqref{intform} and are repeated here for the convenience of the reader,
\beq
  \bal
    &S^2 Y_2 = -2,\quad &&S^2 Y_3 = -1,\quad && S Y_1 Y_2 = 2,&&\quad S Y_1 Y_3 = 1,&&\quad S Y_3^2 = -1,\\
    &Y_1Y_2^2 = 3, \quad &&Y_1Y_2Y_3 = 1, \quad &&Y_2^3 = 3, &&\quad Y_2Y_3^2 = -1.
  \eal
\eeq\\
For the analysis of the D-term supersymmetry condition and the possibility of decoupling gravity we require a basis for the K\"ahler cone. One possible choice for this consists of
\beq
 K_1 = Y_1 + Y_2, \quad K_2 = Y_2, \quad K_3 = X + Y_1, \quad K_4 = Y_1 + Y_3.
\eeq
The corresponding intersection numbers are
\beq \label{KaehlerIntersec}
  \bal
    &K_1^3 = 12, \qquad &&K_1 K_2 K_3 = 5, \qquad &&K_2^3 = 3, \qquad &&K_2 K_4^2 = 1,\\
    &K_1^2 K_2 = 9, \qquad &&K_1 K_2 K_4 = 4, \qquad &&K_2^2 K_3 = 3, \qquad &&K_3^3 = 0,\\
    &K_1^2 K_3 = 7, \qquad &&K_1 K_3^2 = 2, \qquad &&K_2^2 K_4 = 3, \qquad &&K_3^2 K_4 = 1,\\
    &K_1^2 K_4 = 5, \qquad &&K_1 K_3 K_4 = 4, \qquad &&K_2 K_3^2 = 2, \qquad &&K_3 K_4^2 = 1,\\
    &K_1 K_2^2 = 6, \qquad &&K_1 K_4^2 = 1, \qquad &&K_2 K_3 K_4 = 3, \qquad &&K_4^3 = 0.
  \eal
\eeq

\subsubsection*{The GUT divisor as a $dP_5$}
In order to demonstrate that the GUT divisor $y_3=0$ is indeed a $dP_5$-surface, we now explicitly construct it in terms of the coordinates defined in \tableref{exampleData}. We start with the polynomial whose zero-locus defines the base as a divisor in $Y$ in the class $2 Y_1 + Y_2 + Y_3$. It can be evaluated as
\beq
  P_{B} = \sum_{k=0}^1 \sum_{l=0}^{1-k} \sum_{m=0}^{2-k} y_2^k y_7^{1-k} y_3^l y_4^{1-k-l} y_8^m y_5^{2-k-m} P_{k+l+m}(y_1, y_6) . 
\eeq
Now, the GUT-divisor $S$ is defined by $y_3 = 0$, restricting the above polynomial to $l=0$. Explicitly,
\beq
  P_{GUT} = y_7 y_4 ( y_5^2 + y_5 y_8 f_1 + y_8^2 f_2) + y_2 (y_5 g_1 + y_8 g_2),
\eeq
where $f_n$, $g_n$ are generic polynomials of degree $n$ in $y_1$ and $y_6$. Next, we distinguish between the various patches, fixing three coordinates in each case. Due to this fixing, only one scaling relation remains from the initial four:
\beq
  (y_1, y_2, y_3, y_4, y_5, y_6, y_7, y_8) \sim (\sigma y_1, \rho y_2, \lambda y_3, \lambda \sigma y_4, \sigma \xi y_5, \sigma y_6, \frac{\rho}{\lambda \sigma \xi} y_7, \xi y_8) .
\eeq
We give one example of this to illustrate the procedure following the choice of patch. Let us choose $y_2 = y_7 = y_8 = 1$. This leaves $P_{GUT} = y_4 ( y_5^2 + y_5 f_1 + f_2) + (y_5 g_1 + y_8 g_2)$ and the relation $(y_1, y_2, 0, y_4, y_5, y_6, y_7, y_8) \sim (\sigma y_1, y_2, 0, y_4, \sigma y_5, \sigma y_6, y_7, y_8)$. Next, we multiply the polynomial by $y_5^2$ and rewrite it in terms of $(t_0, t_1, t_2, t_3) = (y_5, y_1, y_6, y_4y_5)$:
\beq
  \bal
    P_{GUT} = &\ t_3 (\alpha_{00} t_0^3 + t_0^2 (\alpha_{01} t_1 + \alpha_{02} t_2) + t_0 (\alpha_{11} t_1^2 + \alpha_{12} t_1 t_2 + \alpha_{22} t_2^2)\\
    & + (t_0^3 (\beta_{01} t_1 + \beta_{02} t_2) + t_0^2 (\beta_{11} t_1^2 + \beta_{12} t_1 t_2 + \beta_{22} t_2^2)) , 
  \eal
\eeq
where the Greek letters are simply coefficients. For ease of notation we will drop them in the following.\footnote{One can easily check that the same analysis can be carried out including the coefficients.} Then rearranging gives
\beq
  P_{GUT} =  t_0^3 (t_1 + t_2 + t_3) + t_0^2(t_1^2 + t_1 t_2 + t_2^2 + t_1 t_3 + t_2 t_3) + t_0 t_3 (t_1^2 + t_1 t_2 + t_2^2) .
\eeq
Clearly, each $t_i$ has weight $1$ under the remaining scaling relation and thus the polynomial forms a quartic in $\mathbb{P}_3$. Now, in order to obtain the standard form for the $dP_5$, i.e.\ the intersection of two quadrics in $\mathbb{P}_4$, define a map $\mathbb{P}_3 \rightarrow \mathbb{P}_4$ as follows,
\beq
  (x_0, x_1, x_2, x_3, x_4) = (t_0^2, t_0 t_1, t_0 t_2, t_0 t_3, t_1^2 + t_1 t_2 + t_2^2).
\eeq
Then clearly, the image of the map defines the first quadric,
\beq
  Q = x_1^2 + x_1 x_2 + x_2^2 - x_0 x_4 = 0,
\eeq
while the image of the quartic (the $P_{GUT}$ polynomial) defines the second quadric,
\beq
  P = x_0 (x_1 + x_2 + x_3) + x_1^2 + x_2^2 + x_1 x_2 + x_1 x_3 + x_2 x_3 + x_3 x_4 = 0.
\eeq
Since each $x_i$ has the same charge under the left-over scaling relation, we thus arrive at a $\mathbb{P}_4[2,2]$ representation of the GUT-divisor, which is the representation of a $dP_5$.\\

\subsection{More three-generation flux solutions}
\label{sec_moresol1}

In section \ref{sec_3Gen} we present the gauge flux for a three-generation model consistent with the quantisation conditions. Here we provide more details on the computation.

Let us start with the ansatz
\beq
  x = 4 \lambda,\ \  x \in \mathbb{Z}, \qquad \zeta = -x B|_S + 2 S|_S + 4 \tilde{\zeta},\ \  \zeta \in \mathbb{Z},
\eeq
where $\tilde{\zeta} = (m_0 X + m_1 Y_1 + m_2 Y_2 + m_3 Y_3)|_S$  and  $-B|_S = (S + Y_1 + Y_2)|_S$. Then the requirement $\chi_{5_H} = 0$ leads to
\beq
x = 5 - 10 (m_1 - m_0) - 6 m_2 - 3 m_3,
\eeq
while $\chi_{10} = \pm 3$ implies
\beq
\chi_{10} = -11 + 22 (m_1 - m_0) +12 m_2 + 6 m_3 = \pm3.
\eeq
As one can see immediately, both these expressions only depend on the difference of $m_1$ and $m_0$. This is due to the fact that the class $S + Y_1$ is trivial on $S$, $X|_S = - Y_1|_S$, so that the coefficients of $m_0$ and $m_1$ are of equal magnitude and opposite sign. Thus, for each solution $(m_1-m_0,m_2,m_3)$ we have a redundant infinity of solutions, each of which restricts to the same $\zeta$ on S.\\

As further given in section \ref{sec_3Gen} the value of $\chi/24$ for this model is $226$. Taking into account the  hypercharge flux contribution this allows for a maximal extra flux contribution of $224$ in agreement with the tadpole cancellation condition. This gives an additional constraint on the allowed solutions and the remaining ones are listed in \tableref{zetasolutions}.\\

\begin{table}[ht]
  \centering
  \begin{tabular}{r|r@{\ \;\;}r@{\ \;\;}r@{\ \;\;}|r@{\ \;\;}r@{\ \;\;}|c}
    $N_{flux}$ & $c_1 - c_0$ & $c_2$ & $c_3$ & $x$ & $\chi_{10}$$^\big._\big.$ & $\zeta$\\
    \hline\hline
    $-144$ & $2\ \ \ $ & $-3$ & $0$ & $3$ & $-3$$^\big.$ & $(5 S + 11 Y_1 - 9 Y_2)|_S$\\
    $-144$ & $-1\ \ \ $ & $3$ & $0$ & $-3$ & $3$$_\big.$ & $(-S - 7 Y_1 + 9 Y_2)|_S$\\
    \hline
    $-160$ & $2\ \ \ $ & $-2$ & $-1$ & $0$ & $3$$^\big.$ & $(2 S + 8 Y_1 - 8 Y_2 - 4 Y_3)|_S$\\
    $-160$ & $-1\ \ \ $ & $3$ & $-1$ & $0$ & $-3$ & $(2 S - 4 Y_1 + 12 Y_2 - 4 Y_3)|_S$\\
    $-160$ & $2\ \ \ $ & $-3$ & $1$ & $0$ & $3$ & $(2 S + 8 Y_1 - 12 Y_2 + 4 Y_3)|_S$\\
    $-160$ & $-1\ \ \ $ & $2$ & $1$ & $0$ & $-3$$_\big.$ & $(2 S - 4Y_1 + 8 Y_2 + 4 Y_3)|_S$\\
    \hline
    $-184$ & $2\ \ \ $ & $-2$ & $-2$ & $3$ & $-3$$^\big.$ & $(5 S + 11 Y_1 - 5 Y_2 - 8 Y_3)|_S$\\
    $-184$ & $-1\ \ \ $ & $4$ & $-2$ & $-3$ & $3$ & $(-S - 7 Y_1 + 13 Y_2 - 8 Y_3)|_S$\\
    $-184$ & $2\ \ \ $ & $-4$ & $2$ & $3$ & $-3$ & $(5 S + 11 Y_1 - 13 Y_2 + 8 Y_3)|_S$\\
    $-184$ & $-1\ \ \ $ & $2$ & $2$ & $-3$ & $3$$_\big.$ & $(-S - 7 Y_1 + 5 Y_2 + 8 Y_3)|_S$\\
    \hline
  \end{tabular}
  \caption{\small Solutions for $\zeta$ leading to three-generation SU(5) GUT models in agreement with D3-tadpole cancellation and D-term supersymmetry.}
  \tablelabel{zetasolutions}
\end{table}
 
\section{A further example based on ${\mathbb P}^4[3]$ } \label{appExample2}

In this section we provide another fourfold, obtained again from an elliptic fibration over ${\mathbb P}^4[3]$ by the described blow-up procedure, together with some flux solutions leading to three generation $SU(5)$ GUTs. The toric data along with the corresponding U(1)-charges of the various divisors are given in \tableref{example2Data}.

\begin{table}[ht]
  \centering
  \begin{tabular}{c|r@{\,$=$\,(\,}r@{,\;\;}r@{,\;\;}r@{,\;\;}r@{,\;\;}r@{,\;\;}r@{\,)\;\;}|c|c@{\ \;\;}c@{\ \;\;}c@{\ \;\;}c@{\ \;\;}c@{\ \;\;}c@{\ \;\;}c@{\ \;\;}} 
    nef-part. &\multicolumn{7}{c|}{vertices} & coords. & \multicolumn{7}{c}{U(1)-charges}$_\big.$ \\
    \hline\hline
    & $N_0$ & $0$ & $0$ & $0$ & $0$ & $0$ & $0$ & $x_0$$^\big._\big.$ & $B$ & $S$ & $Y_1$ & $Y_2$ & $Y_3$ & $Y_4$ & $Y_5$\\
    \hline
    $\nabla_1$ & $\nu_1$ & $-1$ & $0$ & $0$ & $0$ & $0$ & $0$ & $y$$^\big.$ & $3$ & $3$ & $3$ & $\cdot$ & $\cdot$ & $3$ & $\cdot$\\
    & $\nu_2$ & $0$ & $-1$ & $0$ & $0$ & $0$ & $0$ & $x$  & $2$ & $2$ & $2$ & $\cdot$ & $\cdot$ & $2$ & $\cdot$\\
    & $\nu_3$ & $3$ & $2$ & $0$ & $0$ & $0$ & $0$ & $z$  & $1$ & $\cdot$ & $\cdot$ & $\cdot$ & $\cdot$ & $\cdot$ & $\cdot$\\
    & $\nu_4$ & $3$ & $2$ & $-1$ & $-1$ & $-1$ & $-1$ & $y_1$  & $\cdot$ & $\cdot$ & $1$ & $\cdot$ & $\cdot$ & $\cdot$ & $\cdot$\\
    & $\nu_5$ & $3$ & $2$ & $1$ & $0$ & $0$ & $0$ & $y_2$ & $\cdot$ & $\cdot$ & $\cdot$ & $\cdot$ & $\cdot$ & $1$ & $\cdot$\\
    & $\nu_6$ & $3$ & $2$ & $-1$ & $-1$ & $0$ & $0$ & $y_3$$_\big.$  & $\cdot$ & $1$ & $\cdot$ & $\cdot$ & $\cdot$ & $\cdot$ & $\cdot$\\
    \hline
    $\nabla_2$ & $\tilde \nu_1$ &   $0$  &   $0$  &   $0$  &   $0$  & $0$ & $1$ & $y_4$$^\big.$ & $\cdot$ & $\cdot$ & $1$ & $\cdot$ & $\cdot$ & $\cdot$ & $\cdot$\\
    & $\nu_7$ &   $0$  &   $0$  &   $0$  &   $0$  &   $1$  &   $0$ & $y_5$ & $\cdot$ & $\cdot$ & $1$ & $1$ & $1$ & $\cdot$ & $1$\\
    & $\nu_8$ &   $0$  &   $0$  &   $0$  &   $1$  &   $0$  &   $0$ & $y_6$ & $\cdot$ & $1$ & $1$ & $\cdot$ & $\cdot$ & $\cdot$ & $\cdot$\\
    & $\nu_9$ &   $0$  &   $0$  &   $0$  &   $0$  &   $-1$  &   $0$ & $y_7$ & $\cdot$ & $\cdot$ & $\cdot$ & $\cdot$ & $\cdot$ & $\cdot$ & $1$\\
    & $\nu_{10}$ &   $0$  &   $0$  &   $-1$  &   $0$  &   $0$ &    $0$ & $y_8$ & $\cdot$ & $-1$ & $-1$ & $-2$ & $-1$ & $1$ & $\cdot$\\
    & $\nu_{11}$ &   $0$  &   $0$  &   $-2$  &   $0$  &   $-1$ &    $0$ & $y_9$ & $\cdot$ & $\cdot$ & $\cdot$ & $1$ & $\cdot$ & $\cdot$ & $\cdot$\\
    & $\nu_{12}$ &   $0$  &   $0$  &   $-1$  &   $0$  &   $-1$ &    $0$ & $y_{10}$$_\big.$ & $\cdot$ & $\cdot$ & $\cdot$ & $\cdot$ & $1$ & $\cdot$ & $\cdot$\\
    \hline
  \end{tabular}
  \caption{\small Toric data for a second fourfold based on $\mathbb P^4[3]$, including the U(1)-charges.}
  \tablelabel{example2Data}
\end{table}

From this, one obtains $c_1(B) = (-B)|_B = (S + Y_1 + Y_4)|_B$, so that $c_1(S) = (Y_1 + Y_4)|_B$. The intersection numbers are as follows
\beq
  \bal
    &S^2 Y_2 = -1, \quad && S Y_2^2 = -1, \quad && Y_1 Y_3 Y_5 = 1, \quad && Y_2^3 = 1, \quad && Y_4^3 = 2,\\
    &S^2 Y_4 = -2, \quad && Y_1 Y_2^2 = -1, \quad && Y_1 Y_4^2 = 2, \quad && Y_2 Y_3^2 = -1, \quad && Y_4^2 Y_5 = 1,\\
    &S Y_1 Y_2 = 1, \quad && Y_1 Y_2 Y_3 = 1, \quad && Y_1 Y_4 Y_5 = 1, \quad && Y_3^3 = 4, \quad && Y_4 Y_5^2 = -1,\\
    &S Y_1 Y_4 = 2, \quad && Y_1 Y_3^2 = -2, \quad && Y_1 Y_5^2 = -1, \quad && Y_3^2 Y_5 = -1, && Y_5^3 = 1.
  \eal
\eeq\\

Now in this case $\chi/24 = 166$, allowing for a maximal flux contribution to the tadpole of $-164$  (in addition to the $-2$ units of hypercharge flux). Starting with the same ansatz as above, only changing $\tilde{\zeta} = (m_0 S + m_1 Y_1 + m_2 Y_2 + m_3 Y_3 + m_4 Y_4 + m_5 Y_5)|_S$ and $-B = S + Y_1 + Y_4$, the conditions on $\chi_{5_H}$ and $\chi_{10}$ result in
\beq
x = 5 - 10 (m_1 - m_0) - 3 m_2 - 6 m_4,
\eeq
and
\beq
\chi_{10} = -11 + 22 (m_1 - m_0) + 6 m_2 + 6 m_4 = \pm3.
\eeq

One notes that both equations are entirely independent of $m_3$ and $m_5$, and furthermore, that mapping $(m_2, m_4)$ to $(m_3, m_2)$ of the previous example leads to the same equations as obtained there. Thus the same analysis as above applies, with in this case a 3-dimensional infinite lattice of redundancy, resulting from the fact that for this model $(S+Y_1)|_S = 0$, $Y_3|_S = 0$, and $Y_5|_S = 0$. The now more restricted set of distinguishable solutions is given in \tableref{zetasolutions2}.

\begin{table}[ht]
  \centering
  \begin{tabular}{r|r@{\ \;\;}r@{\ \;\;}r@{\ \;\;}|r@{\ \;\;}r@{\ \;\;}|c}
    $N_{Flux}$ & $c_1 - c_0$ & $c_2$ & $c_4$ & $x$ & $\chi_{10}$$^\big._\big.$ & $\zeta$\\
    \hline\hline
    $-144$ & $2\ \ \ $ & $0$ & $-3$ & $3$ & $-3$$^\big.$ & $(5 S + 11 Y_1 - 9 Y_4)|_S$\\
    $-144$ & $-1\ \ \ $ & $0$ & $3$ & $-3$ & $3$$_\big.$ & $(-S - 7 Y_1 + 9 Y_4)|_S$\\
    \hline
    $-160$ & $2\ \ \ $ & $-1$ & $-2$ & $0$ & $3$$^\big.$ & $(2 S + 8 Y_1 - 4 Y_2 - 8 Y_4)|_S$\\
    $-160$ & $-1\ \ \ $ & $-1$ & $3$ & $0$ & $-3$ & $(2 S - 4 Y_1 - 4 Y_2 + 12 Y_4)|_S$\\
    $-160$ & $2\ \ \ $ & $1$ & $-3$ & $0$ & $3$ & $(2 S + 8 Y_1 + 4 Y_2 - 12 Y_4)|_S$\\
    $-160$ & $-1\ \ \ $ & $1$ & $2$ & $0$ & $-3$$_\big.$ & $(2 S - 4 Y_1 + 4 Y_2 + 8 Y_4)|_S$\\
    \hline
  \end{tabular}
  \caption{\small Flux solutions for 3-generation models in agreement with D3-tadpole constraints.}
  \tablelabel{zetasolutions2}
\end{table}




\clearpage
\bibliography{rev}  
\bibliographystyle{utphys}


\end{document}